\documentclass[preprint,12pt]{elsarticle}
\usepackage{setspace}
\usepackage[utf8]{inputenc}
\usepackage[english]{babel}
\usepackage{mathtools}
\usepackage{subcaption}
\usepackage{mathrsfs}
\usepackage[pdfencoding=auto,psdextra]{hyperref}
\hypersetup{
  colorlinks = true,
  allcolors = blue,
}
\usepackage{color}
\usepackage{xcolor}
\usepackage{amsthm}
\usepackage{breakcites}
\usepackage[body={7in, 9in},left=1in,right=1in]{geometry}
\usepackage{amsmath,amssymb}
\DeclareMathOperator*{\argmax}{arg\,max}
\DeclareMathOperator*{\argmin}{arg\,min}
\newcommand{\norm}[1]{\left\|#1\right\|}
\usepackage{bm}
\usepackage{bbm}
\usepackage[]{graphicx}
\usepackage[export]{adjustbox}

\usepackage{url}
\usepackage{doi}

\newcommand{\bB}{\mathbf{B}}
\newcommand{\bD}{\mathbf{D}}
\newcommand{\bA}{\mathbf{A}}

\newcommand{\bL}{\mathbf{L}}
\newcommand{\bU}{\mathbf{U}}

\newcommand{\bu}{\mathbf{u}}
\newcommand{\bv}{\mathbf{v}}

\newcommand{\bone}{\textbf{1}}
\newcommand{\bzz}{\textbf{0}}
\newcommand{\bg}{\mathbf{g}}

\newcommand{\bQ}{\mathbf{Q}}

\newcommand{\im}{\text{im}}
\newcommand{\bH}{\mathbf{H}}
\newcommand{\bh}{\mathbf{h}}
\newcommand{\bef}{\mathbf{f}}
\newcommand{\bfh}{\mathbf{\hat{f}}}

\newcommand{\bI}{\mathbf{I}}

\newcommand{\by}{\mathbf{y}}
\newcommand{\bs}{\mathbf{s}}
\newcommand{\br}{\mathbf{r}}
\newcommand{\bp}{\mathbf{p}}
\newcommand{\bw}{\mathbf{w}}
\newcommand{\bc}{\mathbf{c}}
\newcommand{\bW}{\mathbf{W}}

\newcommand{\bY}{\mathbf{Y}}

\newcommand{\bLambda}{\boldsymbol{\Lambda}}
\newcommand{\bTheta}{\boldsymbol{\Theta}}

\newtheorem{theorem}{Theorem}

\theoremstyle{definition}

\theoremstyle{plain}

\theoremstyle{plain}

\theoremstyle{plain}
\newtheorem{lemma}[theorem]{Lemma}

\theoremstyle{definition}
\newtheorem{example}[theorem]{Example}

\theoremstyle{definition}
\newtheorem{remark}[theorem]{Remark}

\theoremstyle{plain}

\theoremstyle{plain}

\theoremstyle{plain}

\theoremstyle{plain}

\journal{Signal Processing}

\begin{document}

\begin{frontmatter}
\title{Signal Processing on Higher-Order Networks: Livin' on the Edge ... and Beyond}
\author{Michael~T. Schaub$^{1}$, Yu Zhu$^{2*}$, Jean-Baptiste Seby$^{3*}$, T. Mitchell Roddenberry$^{2}$, and Santiago Segarra$^{2}$}
\address{$^{1}$ Department of Computer Science, RWTH Aachen University, Germany \\
          $^{2}$ Department of Electrical and Computer Eng., Rice University, Houston, TX, USA \\  
          $^{3}$ Université Paris-Sud, Orsay, France \\  
          $^*$ Equal contribution }
\date{December 2020}

\begin{abstract}
In this tutorial, we provide a didactic treatment of the emerging topic of signal processing on higher-order networks. 
Drawing analogies from discrete and graph signal processing, we introduce the building blocks for processing data on simplicial complexes and hypergraphs, two common higher-order network abstractions that can incorporate polyadic relationships.
We provide brief introductions to simplicial complexes and hypergraphs, with a special emphasis on the concepts needed for the processing of signals supported on these structures.
Specifically, we discuss Fourier analysis, signal denoising, signal interpolation, node embeddings, and nonlinear processing through neural networks, using these two higher-order network models. 
In the context of simplicial complexes, we specifically focus on signal processing using the Hodge Laplacian matrix, a multi-relational operator that leverages the special structure of simplicial complexes and generalizes desirable properties of the Laplacian matrix in graph signal processing.
For hypergraphs, we present both matrix and tensor representations, and discuss the trade-offs in adopting one or the other.
We also highlight limitations and potential research avenues, both to inform practitioners and to motivate the contribution of new researchers to the area.
\end{abstract}

\begin{keyword}
simplicial complex, hypergraph, higher-order network, graph signal processing, node embeddings
\end{keyword}

\end{frontmatter}

\section{Introduction}

Graphs provide a powerful abstraction for systems consisting of (dynamically) interacting entities.
By encoding these entities as nodes and the interaction between them as edges in a graph, we can model a large range of systems in an elegant, conceptually simple framework.
Accordingly, graphs have been used as models in a broad range of application areas~\cite{Newman2018, Easley2010}, including neuroscience~\cite{Sporns2018, medaglia_2017_brain}, urban transportation~\cite{Derible2011}, and social sciences~\cite{Borgatti2009}. 
Many of these applications may be understood in terms of graph signal processing (GSP), which provides a unifying framework for processing data supported on graphs.
In GSP, we model complex data dependencies as the edges of graphs that relate signals on the nodes.
In this way GSP extends and subsumes classical signal processing concepts and tools such as the Fourier transforms, filtering, sampling and reconstruction of signals, and others, to a graph-based setting~\cite{Sandryhaila2013,Shuman2013,Ortega2018}. 

To enable computations with graph-based data, we typically encode the graph structure in an adjacency matrix or its associated (normalized or combinatorial) Laplacian matrix. 
Rather than considering these matrices as a simple table that records pairwise coupling between nodes, it is fruitful to think of these matrices as linear operators that map data from the node space to itself.
By analyzing the properties of these maps -- e.g., their spectral properties -- we can reveal important aspects both about the graphs themselves as well as signals defined on the nodes.
Choosing an appropriate matrix operator associated with the graph structure is thus a key factor in gaining deeper insights about graphs and graph signals.
In GSP, we call such maps that relate data associated with different nodes \emph{graph shift operators}.
Graph shift operators are natural generalizations of the classical time delay, and constitute the fundamental building blocks of graph filters and other more sophisticated processing architectures~\cite{segarra2017optimal}.
The rapid advancement of GSP has benefited significantly from spectral and algebraic graph theory~\cite{Chung1997}, in which the properties of matrices such as the adjacency matrix and the Laplacian have been extensively studied.

By construction, graph-based representations do not account for interactions between more than two nodes, even though such multi-way interactions are widespread in complex systems: multiple neurons can fire at the same time \cite{giusti2016}, biochemical reactions usually include more than two proteins \cite{klamt2009}, and people interact in small groups \cite{kee2013}.
To account for such polyadic interactions, a number of modeling frameworks have been proposed in the literature to represent higher-order relations, including simplicial complexes~\cite{Hatcher2002}, hypergraphs~\cite{Berge1989}, and others~\cite{Frankl1995}.
However, in comparison to this line of work on representing the \emph{structure} of complex multi-relational systems, the literature on the \emph{data processing} for signals defined on higher-order networks is comparatively sparse.
In this tutorial paper, we focus on the topic of \emph{signal processing on simplicial complexes and hypergraphs}.
Following a high-level didactic style, we concentrate on the algebraic representations of these objects, and discuss how the choice of this algebraic representation can influence the way in which we analyze and model signals associated with higher-order networks.

Similarly to graphs, higher-order interactions can be encoded in terms of matrices or, more generally, tensors.
Two of the most prominent abstractions for such polyadic data are simplicial complexes~\cite{Hatcher2002} and hypergraphs~\cite{Berge1989}.
As we will see in the following, both of these abstractions have certain advantages and disadvantages: Hypergraphs are somewhat more flexible in terms of the relationships they can represent, which can be desirable in terms of modeling.
Indeed a simplicial complex may be interpreted as a specific hypergraph for which only certain sets of hyperedges are allowed.
The advantage of simplicial complexes, however, is that this additional structure provides deep links to computational geometry and algebraic topology, which can facilitate both the computation and interpretation of the processed signals~\cite{robinson2014topological}.

Analogously to the graph case, we encode higher-order relations in terms of incidence matrices or tensors that provide an algebraic description of these two data models.
Clearly, the choice of the linear (or multilinear) operator representing higher-order interactions will matter for revealing interesting properties about the data, leading to the key question of how to choose an appropriate abstraction for this kind of data.
In comparison to graphs, the analysis of higher-order interaction data is more challenging due to several factors: 
(i)~There exists a combinatorially large number of possible interactions: two-way, three-way, and so on. Hence, very large matrices and tensors are needed to capture all these relations;
(ii)~The large dimensionality of these representations gives rise to computational and statistical issues on how to efficiently extract information from higher-order data; and
(iii)~The theory on the structure of higher-order networks is largely unexplored relative to that of graphs.
In the following, we will primarily focus on the question of choosing an appropriate algebraic descriptor to implement various signal processing tasks on simplicial complexes and hypergraphs.
Specifically, we will consider the modeling assumptions inherent to an abstraction based on simplicial complexes versus hypergraphs, and discuss the relative advantages and disadvantages of a number of associated matrix and tensor descriptions that have been proposed.
To make our discussions more concrete we provide a number of illustrative examples to demonstrate how the choice of an algebraic description can directly effect the type of results we can obtain.

\noindent\textbf{Outline.} 
We first briefly recap selected  concepts from signal processing and GSP in Section~\ref{section:preliminariesGraphs}. 
In Section~\ref{section:SC}, we present tools from algebraic topology and their use in representing higher-order interactions with simplicial complexes. 
In Section~\ref{section:learning_SC}, we describe methods to analyze signals defined on simplicial complexes.
We then turn our attention to hypergraphs in Section~\ref{section:hypergraphs}, and focus on the modeling of higher-order interactions via hypergraphs. 
Section~\ref{section:learning_hypergraphs} then builds on these models and outlines some of the existing methods for signal processing and learning on hypergraphs. 
Finally, in Section~\ref{section:Discussion}, we close with a brief discussion summarizing the main takeaways and laying out directions for future research.

\section{Signal processing on graphs: A selective  overview}\label{section:preliminariesGraphs}

Before discussing signal processing on higher-order networks, we revisit principles from signal processing and GSP~\cite{Sandryhaila2013,Shuman2013,Ortega2018} and recall some important problem setups, which will later guide our discussion on higher-order signal processing. 
In this tutorial, we focus on undirected graphs (and higher-order networks), although signal processing on directed graphs has been studied as well~\cite{marques2020signal,Furutani2019}. 

\subsection{Central tenets of discrete signal processing}

In discrete signal processing (DSP), signals are processed by filters. 
A linear filter $\bH$ is an operator that takes a signal as input and produces a transformed signal as output.
This linear filtering operation is represented by a matrix-vector multiplication $\bs_{out} = \bH \bs_{in}$ and defines a \emph{linear} system. 
A special role is played by the circular \emph{time shift} filter $\mathbf{S}$, a linear operator that delays the signal by one sample. 
This so-called shift operator underpins the class of time shift-invariant filters, which is arguably the most important class of linear filters in practice.
Specifically, in classical DSP, every linear time shift-invariant filter can be built based on a matrix polynomial of the time-shift $\mathbf{S}$~\cite{oppenheim2009discrete}.

A filter represented by the matrix $\bH$ is shift-invariant if it commutes with the shift operator, i.e., $\mathbf{S} \bH = \bH \mathbf{S}$.
This implies that $\bH$ and $\mathbf{S}$ preserve each others eigenspaces. 
Since the cyclic shift $\mathbf{S}$ is a circulant matrix that is diagonalizable by discrete Fourier modes, this implies that the action of any shift-invariant linear filter in DSP can be understood by means of a Fourier transform. 
Specifically, the eigenvectors of the cyclic time-shift operator provide an orthogonal basis for linear time shift-invariant processing of discrete-time signals.
Thus time-shift invariant filters are naturally interpretable by Fourier analysis~\cite{oppenheim2009discrete}.

\subsection{Graphs, incidence matrices, and the graph Laplacian}
An undirected graph $\mathcal{G}$ is defined by a set of nodes $\mathcal{V} = \{v_1, \cdots, v_N\}$ with cardinality $N$ and a set of edges $\mathcal{E}$ with cardinality $E$ composed of unordered pairs of nodes in $\mathcal{V}$. 
Edges can be stored in the symmetric adjacency matrix $\bA$ whose entries are given by $A_{ij} = A_{ji} = 1$ if $\{i,j\} \in \mathcal{E}$ and $0$ otherwise. 
Given the degree matrix $\bD = \text{diag}(\bA \bone)$, the graph Laplacian associated with $\mathcal{G}$ is given by $\bL = \bD - \bA$.
Alternatively to the adjacency matrix $\bA$, we can collect interactions between the nodes in the graph via the incidence matrix $\bB \in \mathbbm{R}^{N \times E}$. 
For each edge $e$ we define an arbitrary orientation, which we denote by $e= (i,j)$.
We think of such an edge $e$ as being oriented from tail node $i$ to its head node $j$.
Based on this orientation, the incidence matrix $\bB$ is defined such that $B_{ie} = - B_{je} = -1$ and $B_{ke} = 0$ otherwise. 
Using this definition we can provide an equivalent expression for the graph Laplacian as $\bL = \bB\bB^\top$. 
In the remainder of this paper, we choose an edge-orientation induced by the lexicographic ordering of the nodes, i.e., edges will always be oriented such that they point from a node with lower index to a node with higher index.
However, we emphasize that this orientation is arbitrary and is distinct from the notion of a \emph{directed} edge.

\subsection{Graph signal processing}\label{ss:gsp}

\begin{figure}
   \centering
    \includegraphics[width = \textwidth]{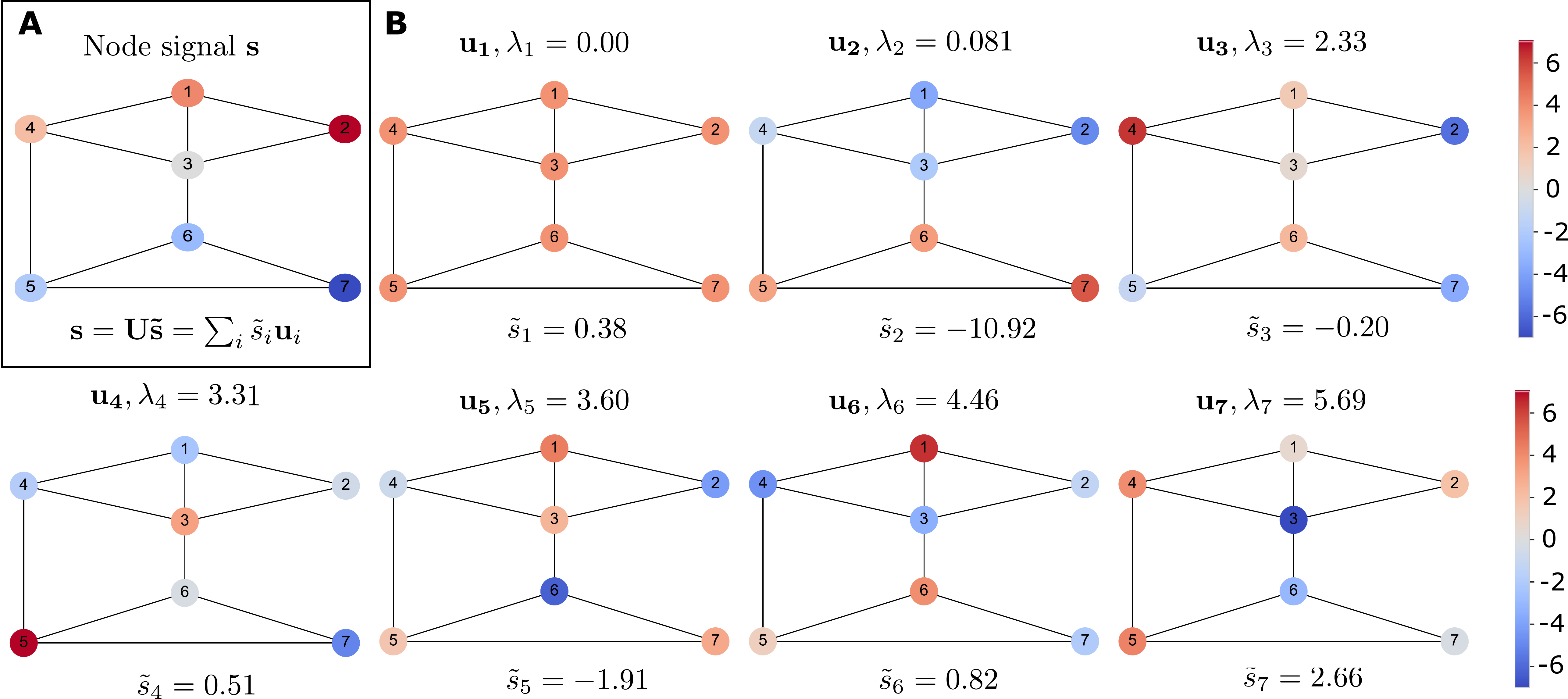}
    \caption{\textbf{Graph signal and its Fourier decomposition.} \textbf{A} Graph signal defined on the nodes of the graph. \textbf{B} Eigenvector and eigenvalue pairs of the graph Laplacian $\bL$. We visualize each of the eigenvectors in terms of a graph signal and order them from low to high graph frequencies, corresponding to a decrease in ``smoothness''. The decomposition of the node signal $\bs$ into this basis provides the Fourier coefficients in $\tilde{\mathbf{s}}$ as indicated at the bottom of each eigenvector representation.}
    \label{fig:SC_example_drawings}
\end{figure}

GSP generalizes the concepts and tools from DSP to signals defined on the nodes of graphs. 
A \emph{graph signal} $\bs : \mathcal{V} \rightarrow \mathbbm{R}$ is a map from the set of nodes $\mathcal{V}$ to the set of real numbers $\mathbbm{R}$.
This defines an isomorphism between the set of nodes and the set of real-valued vectors of length $N$, so any graph signal may be represented as a vector $\bs = [s_1, s_2, \ldots, s_N]^\top \in \mathbbm{R}^N$.
An example of a graph signal can be seen in Figure~\ref{fig:SC_example_drawings}A, where the signal values at each node are indicated by the node color.
Similarly to DSP, filtering in GSP can be represented by a matrix-vector multiplication operation $\bs_{out} = \bH \bs_{in}$.
The analog of the shift operator $\mathbf{S}$ in the GSP setting is any operator that captures the relational dependencies between nodes, including the adjacency matrix $\bA$, the Laplacian matrix $\bL$, or variations of these operators~\cite{Shuman2013, Ortega2018}.

As we are considering undirected graphs here, the choice of a shift operator imparts a natural orthogonal basis $\bU$ in which to represent the signal.
Given the eigenvalue decomposition of the shift operator $\mathbf{S} = \bU \bLambda \bU^\top$ and a filtering weight function $h: \mathbbm{R} \rightarrow \mathbbm{R}$, we can express a shift-invariant filter in this basis as:
\begin{align}
    \bH &= \sum_{k=1}^N h(\lambda_k) \bu_k \bu_k^\top = \bU h(\boldsymbol \Lambda) \bU^\top,
\end{align}
where we have used the shorthand notation $h(\boldsymbol \Lambda) = \text{diag}(h(\lambda_1), \cdots, h(\lambda_N))$. 
By analogy to the Fourier basis in DSP, the eigenvectors $\bU$ of the shift operator are said to define a \emph{graph Fourier transform} (GFT), and $h(\boldsymbol \Lambda)$ is called the frequency response of the filter $\bH$.
Specifically, the GFT of a graph signal $\bs$ is given by $\tilde{\bs} = \bU^\top \bs$, while the inverse GFT is given by $\bs = \bU \tilde{\bs}$ \cite{Ortega2018,Sandryhaila2013}. 

As our discussion emphasizes, any filtered signal $\bs_{out} = \bH \bs_{in}$ on an undirected graph can be understood in terms of three steps: (i) project the signal into the graph Fourier domain, i.e., express it in the orthogonal basis $\bU$ (via multiplication with $\bU^\top$); (ii) amplify certain modes and attenuate others (via multiplication with $h(\boldsymbol \Lambda)$), and (iii) push back the signal to the original node domain (via multiplication with $\bU$).
The choice of an appropriate shift operator is thus crucial, as its eigenvectors define the basis for any shift-invariant graph filter for undirected graphs.
We will encounter this aspect again when considering signal processing on higher-order networks.

In the context of GSP, we focus on the graph Laplacian as a shift operator.
This choice has the following advantages.
First, $\bL$ is positive semidefinite, so that all the graph frequencies (eigenvalues) are real and non-negative.
This enables us to order the GFT basis vectors (eigenvectors) in a natural way.
Second, by considering the variational characterization of the eigenvalues of the Laplacian in terms of the Rayleigh quotient $r(\bs) = \bs^\top \bL \bs / \bs^\top \bs =  \sum_{ij} A_{ij} (s_i- s_j)^2 / (2\|\bs\|^2)$, it can be shown that eigenvectors associated with small eigenvalues have small variation along the edges of the graph (low frequency) and eigenvectors associated with large eigenvalues have large variation along edges (high frequency).
In particular, eigenvectors associated with eigenvalue $0$ are constant over connected components.
An illustration of this is given in Figure~\ref{fig:SC_example_drawings}B, which displays the individual basis vectors of the graph Laplacian, and the coefficients with which these basis vectors would have to be weighted to obtain the previously considered graph signal in Figure~\ref{fig:SC_example_drawings}A.

\subsection{Graph signal processing: Illustrative problems and applications}\label{ss:graph_signal_processing}

Over the last few years, several relevant problems have been addressed using GSP tools including sampling and reconstruction of graph signals~\cite{marques2016sampling, chen2015discrete, anis_2016_sampling}, (blind) deconvolution~\cite{segarra_2016_blind, zhu2020estimating}, and network topology inference~\cite{dong_2016_laplacian,segarra_2017_topo,shen_2017_kernel,mateos_2019_connecting}, to name a few.
We now introduce a subset of illustrative problems and application scenarios that we will revisit in the context of higher-order signal processing.

\subsubsection{Fourier analysis: Node embeddings and Laplacian eigenmaps}\label{subsection:node_Fourier_analysis}

As discussed above, the GFT of a graph signal provides a fundamental tool of GSP.
While we are often interested in filtering a signal and representing it in the vertex space,
the Fourier representation can also be used to gain insight about specific graph components by considering a frequency domain representation of the indicator vector associated with the vertices of interest.
In particular, by considering a truncated Fourier domain representation of the indicator vectors of individual nodes, we can recover a number of spectral node embeddings that have found a broad range of applications (see also~\cite{heimowitz2017unified} for a related discussion).
Specifically, by considering a truncated Fourier domain representation based on the normalized Laplacian as a shift operator, we recover a variant of the so-called Laplacian eigenmaps~\cite{Belkin2003}, and by additionally incorporating a scaling associated with the eigenvalues, we can recover the diffusion map embedding~\cite{coifman2006diffusion,heimowitz2017unified}.

We remark that while most of these spectral node embeddings focus on low frequency eigenvectors, high frequency components can also be of interest for embeddings. 
For instance, if the graph to be analyzed is almost bipartite, then the eigenvectors associated with the highest frequencies of the graph Laplacian will reveal the two (almost) independent node sets in the graph.
Other types of (nonlinear) node embeddings may also be viewed through a GSP lens, e.g., certain node embeddings derived from graph neural networks (cf. Section~\ref{section:gnn-background}).
We refer to~\cite{hamilton2017representation} for an extensive discussion on the highly active area of node representation learning on graphs.

\subsubsection{Signal smoothing and denoising}\label{ss:graph_denoise}

A canonical task in GSP is to denoise (smooth out) a noisy signal $\by = \by_0 + \boldsymbol{\epsilon} \in \mathbb{R}^N$,  where $\by_0$ is the true signal we aim to recover and $\boldsymbol{\epsilon}$ is a vector of zero-mean white Gaussian noise~\cite{chen2014signal}. 
A natural assumption is that the signal should be \emph{smooth} on nearby nodes in terms of the underlying graph, so that neighboring nodes will tend to take on similar values.
Following our above discussion, this amounts to assuming that the signal has a low-pass characteristic, i.e., can be well-represented by the low frequency eigenvectors of the Laplacian. 
Indeed, the eigenvectors of the Laplacian associated with low eigenvalues are smooth on clusters, i.e. their total variation is low within clusters and high over edges between clusters. 

We formalize the above problem in terms of the following optimization problem \cite{Zhou2004, dong_2016_laplacian}
\begin{align}\label{eq:graph_denoising}
    \min_{\hat{\by}} \{ \norm{\hat{\by} - \by }_2^2 + \alpha \hat{\by}^\top \bL \hat{\by}\},
\end{align}
where $\hat{\by}$ is the estimate of the true signal $\by_0$.
The coefficient $\alpha > 0$ can be interpreted as a regularization parameter that trades-off the smoothness promoted by minimizing the quadratic form $\hat{\by}^\top \bL \hat{\by} =  \sum_{ij} A_{ij} (\hat{y}_{i} -  \hat{y}_{j})^2/2$ and the fit to the observed signal in terms of the squared $2$-norm.
The optimal solution for \eqref{eq:graph_denoising} is given by \cite{dong_2016_laplacian}
\begin{align}\label{eq:graph_opt_sl1}
    \hat{\by} = (\bI + \alpha \bL)^{-1} \by.
\end{align}
A different procedure to obtain a signal estimate is the iterative smoothing operation
\begin{align}\label{eq:graph_opt_sl2}
    \hat{\by} = (\bI - \mu \bL)^{k} \by,
\end{align}
for a certain fixed number of iterations $k$ and a suitably chosen update parameter $\mu$.
This may be interpreted in terms of $k$ gradient descent steps of the cost function $\hat{\by}^\top \bL \hat{\by}$.

Matching the signal modeling assumption of a smooth signal, the denoising and smoothing operators defined in \eqref{eq:graph_opt_sl1} and \eqref{eq:graph_opt_sl2} are instances of \emph{low-pass filters}, i.e., filters whose frequency responses $h(\boldsymbol{\lambda}) = \text{diag}(\bU^\top \bH \bU)$ are vectors of non-increasing (decreasing) values. 
In the GSP context, the low-pass filtering operation guarantees that variations over neighboring nodes are smoothed out, in line with the intuition of the optimization problem defined in~\eqref{eq:graph_denoising}.

\subsubsection{Graph signal interpolation}\label{ss:graph_interpolate}

Another common task in GSP is signal interpolation, which can alternatively be interpreted in terms of graph-based \emph{semi-supervised learning}~\cite{chen2015discrete,segarra2016reconstruction}.
Suppose that we are given signal values (labels) for a subset of the nodes $\mathcal{V}^L \subset \mathcal{V}$ of a graph.
Our goal is to interpolate these assignments and to provide a label to all unlabeled nodes $\mathcal{V}^U = \mathcal V \;\backslash\; \mathcal V^L$.

As in the signal denoising case, it is natural to adopt a smoothness assumption that posits that well-connected nodes have similar labels \cite{Chapelle2002}. 
This motivates the following constrained optimization problem \cite{Zhu2003}
\begin{align}\label{eq:graph_SSL}
    &\min_{\hat{\by}} \norm{\bB^\top \hat{\by}}^2_2, \\
    &\text{ s.t. } \hat{y}_i = y_i, \text{ for all } v_i \in \mathcal{V}^L,\nonumber
\end{align}
which aims to minimize the sum-of-squares label difference between connected nodes under the constraint that the observed node labels $y_i$  should be accounted for in the optimal solution.
Notice that the objective function in \eqref{eq:graph_SSL} can again be written in terms of the quadratic form of the graph Laplacian $\norm{\bB^\top \hat{\by}}^2_2= \sum_{(i,j) \in \mathcal{E}} (\hat{y}_i - \hat{y}_j)^2 = \hat{\by}^\top \bL \hat{\by}$, highlighting the low-pass modeling assumption inherent in the optimization problem \eqref{eq:graph_SSL}.

\subsubsection{Graph neural networks}\label{section:gnn-background}

Motivated by spectral interpretations of filters and shift operators in the domain of graph signal processing, \emph{graph neural networks} \cite{bronstein2017geometric,wu2020comprehensive} have emerged as a popular approach to incorporate nonlinearities in the graph signal processing pipeline for purposes of node embedding \cite{cao2016deep,wang2016structural,kipf2016variational}, node classification \cite{kipf2016semi,defferrard2016convolutional}, and graph classification \cite{defferrard2016convolutional}.
Graph neural network architectures combine notions of graph filtering, permutation invariance, and graph Fourier analysis with nonlinear models from the design of neural networks.

One such architecture is the well-known \emph{graph convolutional network}~\cite{kipf2016semi}, which resembles the functional form of \eqref{eq:graph_opt_sl2} with interleaved nonlinear, elementwise activation functions, i.e., for a set of $F_0$ input features gathered in the columns of a matrix $\bY_0\in\mathbbm{R}^{N\times F_0}$,
\begin{equation}\label{eq:basic-gcn}
    \bY_k = \sigma(\bH \bY_{k-1} \bW_k),
\end{equation}
where we take $\bY_K$ for some integer $K$ as the output, $\{\bW_k\in\mathbbm{R}^{F_{k-1}\times F_k}\}_{k=1}^K$ are learnable weight matrices that perform linear transformations in the feature space, $\bH$ is a certain graph filter, and $\sigma(\cdot)$ is a general nonlinear activation function applied elementwise.
Specifically, \cite{kipf2016semi} uses a normalized version of the graph Laplacian as a first-order filter $\bH$, and the ReLU activation function for~$\sigma(\cdot)$.

A closer look at \eqref{eq:basic-gcn} reveals a connection with the iterative smoothing method of \eqref{eq:graph_opt_sl2}.
Taking $\sigma(\cdot)$ to be the identity mapping, we see that \eqref{eq:basic-gcn} can be expressed as a linear graph filter independently applied to each of the $F_0$ features, with output defined as linear combinations of these filtered features at each node via the matrices $\{\bW_k\}$.
That is,
\begin{equation}\label{eq:basic-linear-gcn}
    \bY_K = (\bH^K\bY_0)(\bW_1\bW_2\ldots\bW_K),
\end{equation}
where $\bH^K$ itself represents a shift-invariant graph filter, due to the assumed shift-invariance of $\bH$.
Taking $F_0=F_K=1$ and $\bH=(\bI-\mu\bL)$ recovers the iterative smoothing procedure of \eqref{eq:graph_opt_sl2}.
However, by interleaving nonlinear functions as in \eqref{eq:basic-gcn} and taking linear combinations of features via $\{\bW_k\}$, we allow the architecture to \emph{learn} more sophisticated, nonlinear relationships between the nodes and node features by finding optimal weights $\{\bW_k\}$ for a suitable loss function.

There are many variants of the graph neural network architecture, designed for tasks ranging from semi-supervised learning~\cite{kipf2016semi} to graph classification~\cite{defferrard2016convolutional}.
We refer the reader to the survey paper~\cite{wu2020comprehensive} for further details, as well as~\cite{gama2018convolutional} for a view focused on graph signal processing in particular.

\section{Modeling higher-order interactions with simplicial complexes}\label{section:SC}

\begin{figure}
    \centering
    \includegraphics[width=\textwidth ]{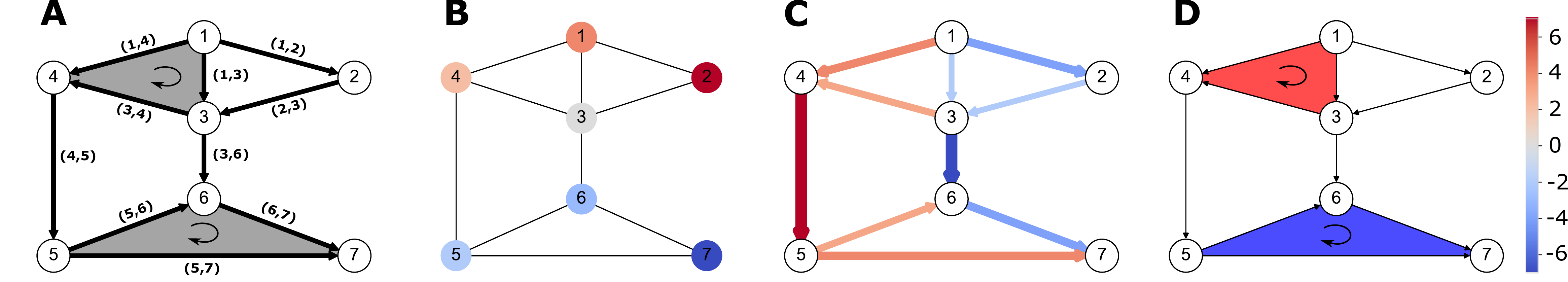}
    \caption{\textbf{Signals on simplicial complexes of different order.} \textbf{A}: Structure of the simplicial complexes used as a running example in the text. Arrows represent the chosen reference orientation. Shaded areas correspond to the $2$-simplices $\{1,3,4\}$ and $\{5,6,7\}$.
    \textbf{B}: Signal on $0$-simplices (nodes).
    \textbf{C}: Signal on $1$-simplices (edges). 
    \textbf{D}: Signal on $2$-simplices (triangles).}
    \label{fig:SC_example}
\end{figure}

In this section, we recap some of the mathematical underpinnings of simplicial complexes.
We focus in particular on the Hodge Laplacian~\cite{Hatcher2002,Lim2015,Schaub2018}, which extends the graph Laplacian as a natural shift operator for simplicial complexes.
Specifically, we discuss how the eigenvectors of the Hodge Laplacian provide an interpretable orthogonal basis for signals defined on simplicial complexes by means of the Hodge decomposition.

\subsection{Background on simplicial complexes}

Given a finite set of vertices $\mathcal{V}$, a $k$-simplex $\mathcal{S}^k$ is a subset of $\mathcal{V}$ with cardinality $k+1$. A simplicial complex $\mathcal{X}$ is a set of simplices such that for any $k$-simplex $\mathcal{S}^k$ in $\mathcal{X}$, any subset of  $\mathcal{S}^k$ must also be in $\mathcal{X}$.
A \emph{face} of a simplex $\mathcal{S}^k$ is a subset of $\mathcal{S}^k$ with cardinality $k$. A \emph{co-face} $\mathcal{S}^{k+1}$ of a simplex $\mathcal{S}^k$ is a $(k+1)$-simplex such that $\mathcal{S}^k$ is a subset of $\mathcal{S}^{k+1}$. 
More detailed discussions and definitions can, e.g., be found in \cite{Grady2010,Lim2015,Munkres2000}.

\begin{example}
Figure~\ref{fig:SC_example}A provides an example of a simplicial complex.
Here, simplices of order $0$ are depicted as nodes, simplices of order $1$ as edges, and simplices of order 2 are displayed as gray, filled triangles. 
Note how the edges $\{1,3\}, \{1,4\}$ and $\{3,4\}$ are faces of the $2$-simplex $\{1,3,4\}$. 
The $2$-simplex $\{5,6,7\}$ is a co-face of the edges $\{5,6\}, \{5,7\}$ and $\{6,7\}$.
\end{example}

For computational purposes, we define an orientation for each simplex by fixing an ordering of its vertices.
This ordering induces a reference orientation by increasing vertex label. 
Based on the reference orientation for each simplex, we introduce a book-keeping of the relationships between $(k-1)$-simplices and $k$-simplices via linear maps called \emph{boundary operators} that record higher-order interactions in networks. 
As the simplicial complexes we consider are all of finite order, these boundary operators can be represented by matrices $\bB_k$.
The rows of $\bB_k$ are indexed by $(k-1)$-simplices and the columns of $\bB_k$ are indexed by $k$-simplices. 
For instance, $\bB_1$ is nothing but the node-to-edge incidence matrix denoted $\bB$ in Section \ref{section:preliminariesGraphs}, while $\bB_2$ is the edge-to-triangle incidence matrix.
\begin{example}
We adopt the lexicographic order to define the reference orientation of simplices in Figure~\ref{fig:SC_example}. The corresponding boundary maps $\mathbf{B}_1$ and $\mathbf{B}_2$ are then given by

\begin{minipage}[t]{0.58 \linewidth}
{\scriptsize
\let\quad\thinspace 
\[ \mathbf{B}_1 = \bordermatrix{
& (1,2)&(1,3)&(1,4)&(2,3)&(3,4)&(3,6)&(4,5)&(5,6)&(5,7)&(6,7) \cr
     1&-1&-1&-1&0&0&0&0&0&0&0\cr
     2&1&0&0&-1&0&0&0&0&0&0\cr
     3&0&1&0&1&-1&-1&0&0&0&0\cr
     4&0&0&1&0&1&0&-1&0&0&0\cr
     5&0&0&0&0&0&0&1&-1&-1&0\cr
     6&0&0&0&0&0&1&0&1&0&-1\cr
     7&0&0&0&0&0&0&0&0&1&1}
\]}
\end{minipage}
\begin{minipage}[t]{0.39\linewidth}
{\scriptsize
\let\quad\thinspace 
\[ \mathbf{B}_2 = \bordermatrix{
& (1,3,4)&(5,6,7)\cr
(1,2)&0&0\cr
(1,3)&1&0\cr
(1,4)&-1&0\cr
(2,3)&0&0\cr
(3,4)&1&0\cr
(3,6)&0&0\cr
(4,5)&0&0\cr
(5,6)&0&1\cr
(5,7)&0&-1\cr
(6,7)&0&1
}
\]}
\end{minipage}
\end{example}

We may consider signals defined on any $k$-simplices (nodes, edges, triangles, etc.) of a simplicial complex as illustrated in Figure~
\ref{fig:SC_example}B-D.
Just like for graph signals, we need to establish an appropriate shift operator to process such signals.
While there are many possibilities, we will show in the next section that a natural choice for the shift operator is the Hodge Laplacian, a generalization of the graph Laplacian rooted in algebraic topology.

\subsection{The Hodge Laplacian as a shift operator for simplicial complexes}\label{subsection:Hodge_Laplacian}

Based on the incidence matrices defined above, we can define a sequence of so-called \emph{Hodge Laplacians}~\cite{Lim2015}.
Specifically, the \emph{$k$-th combinatorial Hodge Laplacian}, originally introduced in~\cite{Eckmann1944}, is given by~\cite{Eckmann1944,Lim2015}: 
\begin{equation}
        \bL_k = \bB_k^\top \bB_k + \bB_{k+1} \bB^\top_{k+1}.
\end{equation}
Notice that, according to this definition, the graph Laplacian corresponds to $\bL_0 = \bB_1 \bB_1^\top$ with $\bB_0 = 0$.
More generally, by equipping all spaces with an inner product induced by positive diagonal matrices, we can define a weighted version of the Hodge Laplacian (see, e.g., \cite{Grady2010,Lim2015,Schaub2018,BensonE11221}). 
This weighted Hodge Laplacian encapsulates operators such as the random walk graph Laplacian or the normalized graph Laplacian as special cases.
For simplicity, in this paper we concentrate on the unweighted case.

Just like the graph Laplacian provides a useful choice for a shift operator for node signals defined on a graph due to its (spectral) properties, the Hodge Laplacian and its weighted variants provide a natural shift operator for signals defined on the edges of a simplicial complex (or graph).
As the edges in our simplicial complexes are equipped with a chosen reference orientation, the Hodge Laplacian is in particular relevant as shift operator if the signals considered are indeed oriented, e.g., correspond to some kind of edge-flow in case of a signal on edges.

Similar to the graph Laplacian, the Hodge Laplacian is positive semi-definite, which ensures that we can interpret its eigenvalues in terms of non-negative frequencies.
Moreover, these frequencies are again aligned with a specific type of signal-smoothness displayed by the eigenvectors of the Hodge Laplacian.
For signals on general $k$-simplices, this notion of smoothness can be understood by means of the so called $\emph{Hodge decomposition}$~\cite{Lim2015,Grady2010,Schaub2018}, which states that the space of $k$-simplex signals can be decomposed into three orthogonal subspaces
\begin{align}\label{eq:Hodge_decomposition}
    \mathbb{R}^{N_k} =  \im(\bB_{k+1})\oplus \im(\bB_k^\top) \oplus \ker(\bL_k),
\end{align}
where $\im(\cdot)$ and $\ker(\cdot)$ are shorthand for the \emph{image} and \emph{kernel} spaces of the respective matrices, $\oplus$ represents the union of orthogonal subspaces, and $N_k$ is the cardinality of the space of signals on $k$-simplices (i.e., $N_0=N$ for the node signals, and $N_1= E$ for edge signals).
Here we have (i) made use of the fact that a signal on a finite dimensional set of $N_k$ simplices is isomorphic to $\mathbbm{R}^{N_k}$; and (ii) implicitly assumed that we are only interested in real-valued signals and thus a Hodge decomposition for a real valued vector space (see~\cite{Lim2015} for a more detailed discussion).

\begin{figure}
    \centering
    \includegraphics[width = \textwidth]{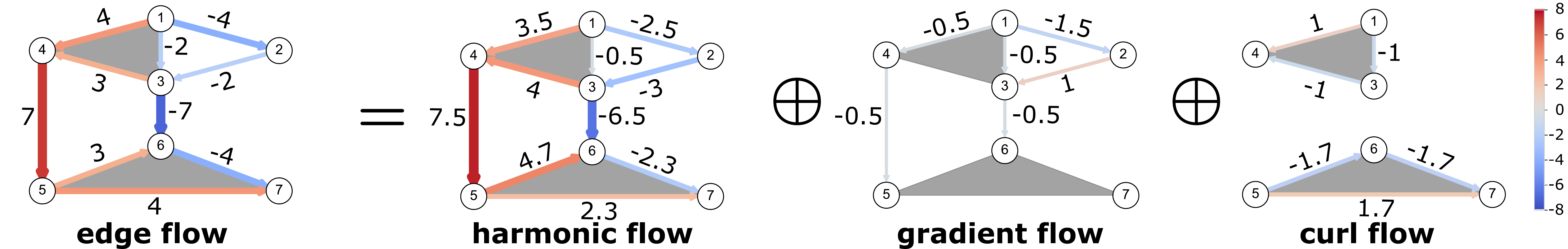}
    \caption{\textbf{Hodge decomposition of the edge flow in the example from Figure \ref{fig:SC_example}}. Any edge flow (left) can be decomposed into a harmonic flow, a gradient flow and a curl flow.}
    \label{fig:Hodge_decomposition}
\end{figure}

To facilitate the discussion on how the Hodge decomposition~\eqref{eq:Hodge_decomposition} can be related to a notion of smooth signals let us consider the concrete case $k = 1$ with Hodge Laplacian $\bL_1 = \bB_1^\top \bB_1 + \bB_2 \bB_2^\top$ for illustration~\cite{Schaub2018,Schaub2018a,Jia2019}.
In this case, we can provide the following meaning to the three subspaces considered in~\eqref{eq:Hodge_decomposition}.
First, the space $\im(\bB_1^\top)$ can be considered as the space of gradient flows (or potential flows).
Specifically, since $\im(\bB_1^\top) = \{\bef = \bB_1^\top \mathbf{v}, \text{for some } \mathbf{v} \in \mathbbm{R}^N\}$ we may create any such flow according to the following recipe: (i) assign a scalar potential to all the nodes; (ii) induce a flow along the edges by considering the difference of the potentials on the respective endpoints.
Clearly, we cannot create a positive net-flow along any closed path within a complex if the flow at every edge is computed according to the gradient (difference) of the node potentials in the chosen reference orientation: the difference between the potentials along any closed path has to sum to zero, by construction.
Accordingly, the space $\ker(\bB_1) = \im(\bB_2) \oplus \ker(\bL_1)$ that is orthogonal to $\im(\bB_1^\top)$ is the so-called cycle space.
As indicated, the cycle space is spanned by two types of cyclic flows.
The space $\im(\bB_{2})$ consists of curl flows and its elements are flows that can be composed of combinations of local circulations along any $2$-simplex.
Specifically, we may assign a scalar potential to each $2$-simplex and consider the induced flows $\bef = \bB_2 \mathbf{t}$, where $\mathbf{t}$ is the vector of $2$-simplex potentials.
Note that every column of $\bB_2$ creates a triangular circulation around the respective $2$-simplex along its chosen reference orientation.
Hence, these flows correspond to local cycles associated with the $2$-simplices present in the simplicial complex.
Finally $\ker(\bL_1)$ is the harmonic space, whose elements correspond to (global) circulations that are not representable as a linear combination of curl flows. 

\begin{example}
In Figure \ref{fig:Hodge_decomposition}, we consider the edge flow $\bc = [-4,-2,4,-2,3,-7,7,3,4,-4]^\top$. The Hodge decomposition $ \bc = \bh \oplus \bg \oplus \br$ enables us to decompose the edge flow $\bc$ into a harmonic, gradient and curl part,  respectively denoted by $\bh, \bg$ and $\br$. Components $\bg$ and $\br$ are given by
\begin{align}
    \bg = \bB_1^\top \bp, \quad \quad\br = \bB_2 \bw.
\end{align}
Since the Hodge decomposition is orthogonal, $\bp$ and $\bw$ are the solutions of the following least squares problems
\begin{align}
    \min_{\bp} \norm{\bB_1^\top \bp - \bc}_2, \quad  \quad \min_{\bw} \norm{\bB_2 \bw - \bc}_2.
\end{align}
The harmonic component satisfies $\bL_1 \bh = \bzz$, and by the orthogonality of the Hodge decomposition, it can be obtained by $\bh = \bc - \bg - \br$.
As explained in the text, $\bg$ is an element of the space $\im(\bB_1^\top)$, i.e., the gradient space or space of cycle-free flows. 
Components $\bh \in \ker(\bL_1)$ and $\br \in \im(\bB_2)$ are elements of the cycle space $\ker(\bB_1) = \im(\bB_2) \oplus \ker(\bL_1)$. As can be seen in Figure \ref{fig:Hodge_decomposition}, the curl component $\br$ can be decomposed into two local circulations, of absolute magnitude $1$ and $1.7$, respectively.
\end{example}

Importantly the gradient, curl and harmonic subspaces are spanned by certain subsets of eigenvectors of $\bL_1$ as the following lemma, which can be verified by direct computation \cite{Barbarossa2020,Schaub2018}, shows.
\begin{lemma}
    Let $\bL_1 = \bB_1^\top \bB_1 + \bB_2 \bB_2^\top$ be the Hodge 1-Laplacian of a simplicial complex.
    Then the eigenvectors associated with nonzero eigenvalues of $\bL_1$ comprise two groups that span the gradient space and the curl space respectively.
    \begin{itemize}
        \item Consider any eigenvector $\bv_i$ of the graph Laplacian $\bL_0$ associated with a nonzero eigenvalue $\lambda_i$. 
            Then $\bu_\text{grad}^{(i)} = \bB_1^\top\bv_i$ is an eigenvector of $\bL_1$ with the same eigenvalue $\lambda_i$. 
            Moreover $\bU_\text{grad} = [\bu_\text{grad}^{(1)},\bu_\text{grad}^{(2)},\ldots]$ spans the space of all gradient flows.
        \item Consider any eigenvector $\mathbf{t}_i$ of the ``2-simplex coupling matrix'' $\mathbf{T} = \bB_2^\top\bB_2$ associated with a nonzero eigenvalue $\theta_i$. 
            Then $\bu_\text{curl}^{(i)} = \bB_2\mathbf{t}_i$ is an eigenvector of $\bL_1$ with the same eigenvalue $\theta_i$. 
            Moreover $\bU_\text{curl} = [\bu_\text{curl}^{(1)},\bu_\text{curl}^{(2)},\ldots]$ spans the space of all curl flows.
    \end{itemize}
\end{lemma}
The above result shows that, unlike for node signals, edge-flow signals can have a high frequency contribution, reflected by a high component in the corresponding projected space, due to two different types of (orthogonal) basis components being present in the signal: a high frequency may arise both due to a curl component as well as a strong gradient component present in the edge-flow.
This has certain consequences for the filtering of edge signals that we will discuss in more detail in the following section.

\section{Signal processing and learning on simplicial complexes}\label{section:learning_SC}

Using the algebraic framework of simplicial complexes as discussed in Section~\ref{section:SC}, in this section we revisit the four signal processing setups considered in Section~\ref{ss:graph_signal_processing}---Fourier analysis and embeddings, smoothing and denoising, signal interpolation, and nonlinear (graph) neural networks---and discuss how these can be extended to simplicial complexes by means of the Hodge Laplacian and associated boundary maps.
For concreteness, we concentrate primarily on edge signals, though the results presented here can be extended to signals on any type of simplices.

\subsection{Fourier analysis: Edge-flow and trajectory embeddings}\label{ssec:fourier_edge_flows}
In the same way that the (normalized) graph Laplacian provides a node embedding of the graph, the eigenvectors of the Hodge Laplacian $\bL_1$ can be used to induce a low-frequency edge embedding. 
As a concrete example, let us consider the harmonic embedding, i.e., the projection of an edge signal $\bef$ into the harmonic subspace, corresponding to signal with zero frequency
\begin{align}\label{eq:harm_projection}
    \mathbf{f}_\text{emb} = \bU_\text{harm}^\top \bef,
\end{align}
where $\bU_\text{harm} = [\bu_\text{harm}^{(1)}, \bu_\text{harm}^{(2)}, \ldots]$ corresponds to eigenvectors of the Hodge Laplacian $\bL_1$ associated with zero eigenvalues.
As explained in Section \ref{section:SC}, the harmonic space spanned by the vectors $\bU_\text{harm}$ corresponds to (globally) cyclic flows that cannot be composed from locally cyclic flows (curl flows).
Analogously to the embedding of nodes via indicator signals projected onto the low frequency eigenvectors (i.e., eigenvectors associated with low eigenvalues) of the graph Laplacian, we can construct embeddings of individual edges using \eqref{eq:harm_projection}.
Unlike for graphs where such node embeddings can indicate a clustering of the nodes \cite{Luxburg2007}, an edge embedding into the harmonic subspace characterizes the position of an edge relative to the harmonic flows.
Since the harmonic flows are in one-to-one correspondence with the $1$-homology of the simplicial complex, i.e., the ``holes'' in the complex that are not filled with faces, such an embedding may be used to identify edges whose location is in accordance with particular harmonic cycles~\cite{Schaub2018,ebli2019notion}.
However, as the edges are equipped with an arbitrary reference orientation, the sign of the projection into the harmonic space is arbitrary.
This is a consequence of the fact that, unlike the graph Laplacian, the Hodge Laplacian is in general not invariant, but equivariant under a change of the reference orientation of the edges (cf. section~\ref{section:GNN}).
To account for this fact, one may use a clustering approach that is invariant to this arbitrary choice of sign.
For instance, we can use subspace clustering as in~\cite{ebli2019notion}, or consider the absolute value of the projection as discussed in~\cite{Schaub2018}.

Rather than aiming at grouping edges together into clusters according to their relative position with respect to the $1$-homology~\cite{ebli2019notion}, we may be interested in grouping sequences of edges corresponding to trajectories on a simplicial complex by projecting appropriate signal indicator vectors of such trajectories into the harmonic space~\cite{Schaub2018}.
Here we represent a trajectory by a vector $\mathbf{f}$ with entries $\mathbf{f}_{(i,j)} = 1$ if the edge $(i,j)$ is part of the trajectory and traversed along the chosen reference orientation,  $\mathbf{f}_{(i,j)} = -1$ if the edge $(i,j)$ is part of the trajectory and traversed opposite to the chosen reference orientation, and $\mathbf{f}_{(i,j)} = 0$ otherwise.

\begin{example}\label{ex:edge_embedding}
In Figure~\ref{fig:embedding_trajectories}A, we construct a simplicial complex by drawing $400$ random points in the unit square and generating a triangular lattice by Delaunay triangulation. We eliminate two points and all their adjacent edges in order to create two ``holes" in the simplicial complex, which are not covered by a $2$-simplex.
These two holes are represented by orange shaded areas and can be interpreted as obstacles through which trajectories cannot pass. 
All (other) triangles are considered as 2-simplices. 
Accordingly, the Hodge Laplacian has two zero eigenvalues associated to two harmonic functions $\mathbf{u}_{\text{harm}}^{(1)}$ and $\mathbf{u}_{\text{harm}}^{(2)}$. 

On the edges of the simplicial complex, we define five trajectories as displayed in Figure~~\ref{fig:embedding_trajectories}A. 
Figure~\ref{fig:embedding_trajectories}B shows the corresponding embeddings of the flow vectors of each trajectory and their evolution in the embedding space. 
More explicitly, for a given trajectory we build the embedding sequentially as follows.
The embedding starts at zero.
We then iteratively project the next edge in the trajectory (accounting for the chosen reference direction) into the harmonic space.
In our case each edge is described by a position $(u_1,u_2)$ in the harmonic space: one component along $\mathbf{u}_{\text{harm}}^{(1)}$ and the other along $\mathbf{u}_{\text{harm}}^{(2)}$. 
The embedding of the trajectory is then obtained from adding these position vectors of the individual edges. 
Note that due to the linearity of the projection operation, this leads to the same final embedding (marked by a red dot) as if we had directly projected the full trajectory vector.

Importantly, the embedding differentiates the topological properties of the trajectories. The magenta and olive green trajectories have a similar embedding since they both pass above the top left obstacle. 
The maroon and green trajectories pass between the two obstacle and have a similar embedding (negative coordinate along $\mathbf{u}_{\text{harm}}^{(1)}$ and zero component along $\mathbf{u}_{\text{harm}}^{(2)}$). 
The orange trajectory is the only one that goes through the right of the bottom right obstacle. 
Hence, its embedding stands out from the other four trajectories in the embedding space.
For a more extensive discussion of these aspects see~\cite{Schaub2018}.

\begin{figure}
    \centering
    \includegraphics[width = \textwidth]{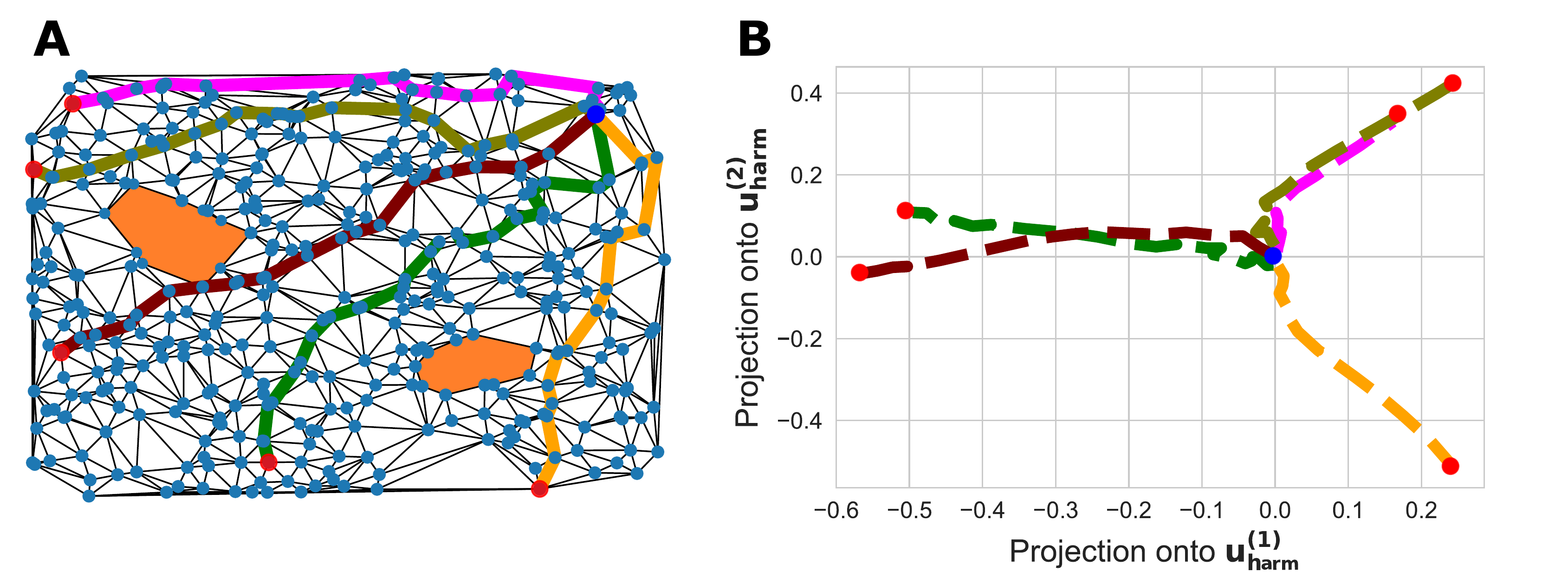}
    \caption{\textbf{Embedding of trajectories defined on a simplicial complex.} \textbf{A} Five trajectories defined on a simplicial complex containing two obstacles, indicated by orange color. The simplicial complex is constructed by creating a triangular lattice from a random set of points and then introducing two ``holes" in this lattice. All triangles in the lattices are assumed to correspond to $2$-simplices.
    \textbf{B} The projection of the trajectories displayed in \textbf{A} into the two dimensional harmonic space of the simplicial complex. 
    Notice that the trajectories that move around the obstacles in a topologically similar way have a similar embedding~\cite{Schaub2018}.}  
    \label{fig:embedding_trajectories}
\end{figure}
\end{example}

As we have seen in the above example, trajectories that behave similarly with respect to the $1$-homology (``holes'') of a simplicial complex will have a similar embedding~\cite{Schaub2018}.
One may thus, for instance, also identify topologically similar trajectories on the simplicial complex by clustering the resulting points in the harmonic embedding.
Such an approach is of interest for a number of applications: One can construct simplicial complexes and appropriate trajectory embedding from a variety of flow data, including physical flows such as buoys drifting in the ocean~\cite{Schaub2018}, or ``virtual'' flows such as click streams or flows of goods and money.
Related ideas for analyzing trajectories have also been considered in the context of traffic prediction~\cite{ghosh2018topological}.

While we have considered here only harmonic embeddings corresponding to signals with zero frequency, other type of embeddings may be of interest as well.
We may, for instance, be interested in gradient-flow-based embeddings, which can be used to define a form of ranking of the nodes in terms of the associated potentials~\cite{Jiang2011}, or be interested in other forms of flows, which are only approximately harmonic~\cite{Jia2019}.

\subsection{Flow smoothing and denoising}
We now revisit the question of smoothing and denoising from the perspective of signals defined in the edge space of a simplicial complex $\mathcal X$.
In parallel, we provide a more in-depth discussion on the basis vectors and notion of a smooth signal encapsulated in the Hodge 1-Laplacian $\bL_1$ and how it differs from the graph Laplacian~\cite{Lim2015, Ortega2018, Barbarossa2018}.

Let us assume that the simplicial complex $\mathcal X$ is associated with oriented flows $\bef^0 \in \mathbb{R}^E$ defined on edges.
Like in the node-based setup discussed in Section~\ref{ss:graph_denoise}, we assume that we can only observe a noisy version $\bef = \bef^0 + \boldsymbol{\epsilon}$ of the true underlying signal, where $\boldsymbol{\epsilon}$ is again a zero-mean white Gaussian noise vector of appropriate dimension. 
By analogy with the graph case, in order to get a smooth estimate $\bfh$ of the true signal $\bef^0$ from the noisy signal $\bef$, it is tempting to adopt the successful procedures from GSP (cf. equation \eqref{eq:graph_denoising}) and solve the following optimization program for the edge-flows $\bef$
\begin{align}\label{eq:edge_denoising}
    \min_{\hat{\bef}} \left\{ \norm{\bfh - \bef }_2^2 + \alpha \bfh^\top \bQ \bfh \right\},
\end{align}
with optimal solution $\bfh = \bH_Q\bef := (\bI + \alpha\bQ)^{-1}\bef$, where the matrix $\bQ$ is a regularizer that needs to be chosen to ensure a smooth estimate.
Following our discussion above, since the filter $\bH_Q$ will inherit the eigenvectors of the regularizer $\bQ$, a natural choice for a regularizer is an appropriate (simplicial) shift operator.

We discuss three possible choices for the regularizer (shift operator) $\bQ$: (i) the graph Laplacian $\bL_\text{LG}$ of the line-graph of the underlying graph skeleton of the complex $\mathcal X$, i.e., the line-graph of the graph induced by the $0$-simplices (nodes) and $1$-simplices (edges) of $\mathcal X$; (ii) the edge Laplacian $\bL_e= \bB_1^\top \bB_1$,  i.e., a form of the Hodge Laplacian that ignores all $2$-simplices in the complex $\mathcal X$ such that $\bB_2=0$; (iii) the Hodge Laplacian $\bL_1 = \bB_1^\top \bB_1 + \bB_2 \bB_2^\top$ that takes into account all the triangles of $\mathcal X$ as well.
Before embarking on this discussion, however, let us illustrate the effects of these choices by means of the following concrete example.

\begin{example}\label{ex:edge_smoothing}
    Figure \ref{fig:flow_smoothing}A displays a conservative (cyclic) flow on a simplicial complex, i.e., all of the flow entering a particular node exits the node again.
This flow is then distorted by a Gaussian noise vector $\boldsymbol{\epsilon}$ in Figure \ref{fig:flow_smoothing}B. 
The estimation error produced by the filter based on the line-graph (Figure \ref{fig:flow_smoothing}C) is comparatively worse (36.54 vs. 1.95 and 1.02 respectively) than the estimation performance of the edge Laplacian (Figure \ref{fig:flow_smoothing}D) and the Hodge Laplacian (Figure \ref{fig:flow_smoothing}E) filters.
\begin{figure}
    \centering
    \includegraphics[width = \textwidth]{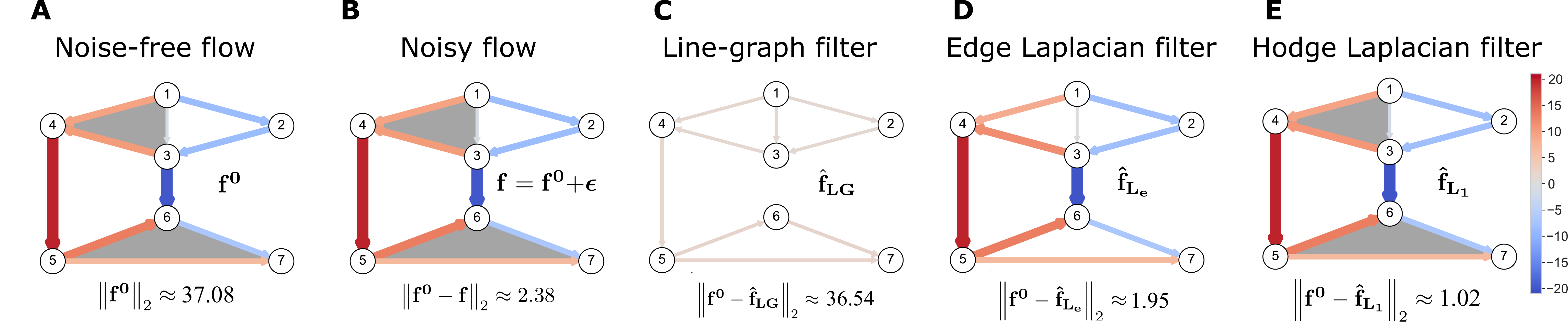}
    \caption{\textbf{Flow smoothing on a graph.} \textbf{A} An undirected graph with a pre-defined and oriented flow $\bef^0$. \textbf{B} The observed flow is a noisy version of the flow $\bef^0$, i.e., $\bef^0$ is distorted by a Gaussian white noise vector~$\boldsymbol{\epsilon}$. \textbf{C} We denoise the flow by applying a Laplacian filter based  on the line-graph. This filter performs worse compared to the edge space filters in \textbf{D} and \textbf{E} that account for flow conservation. \textbf{D} Denoised flow obtained after applying the filter based on the edge Laplacian. \textbf{E} Denoised flow obtained after applying the filter based on the Hodge Laplacian. The estimation error is lower than in the edge Laplacian case as the filter accounts for filled faces in the graph.}
    \label{fig:flow_smoothing}
\end{figure}
\end{example}

Let us explain the results obtained from the individual filters in the above example in more detail, starting with the line-graph approach.
As can be seen from Figure~\ref{fig:flow_smoothing}C, in this case the filtering operation leads to an increased error compared to the noisy input signal.
This ineffective filtering result by means of the line-graph Laplacian has been observed in~\cite{Schaub2018a}. 
The reason for this unintended behavior is that the line-graph Laplacian is not well-suited as a shift operator for flow signals.
The basis functions given by the eigenvectors of the line-graph Laplacians induce a notion of smooth, low frequency signals that supposes that signals on adjacent edges in the simplicial complex have a small difference.
This is equivalent to the fact that low-frequency modes in the node space do not vary a lot on tightly connected nodes on a graph.
However, for flow signals this type of smoothness induced by eigenvectors of the line-graph Laplacian as shift operator is often not appropriate.
Specifically, real-world flow signals typically display a large degree of flow conservation: most of the flow signal entering a node exits the node again, but the relative allocation of the flow to the edges does not have to be similar.
Moreover, the line-graph Laplacian does not reflect the arbitrary orientation of the edges, so that performance is dependent on the chosen sign of the flow.
Notice, however, that the line-graph \emph{can} be a valid representation to process signals on edges that are not encoding flows and, as such, do not have a natural orientation. For example, one might expect the level of congestion on different roads to vary smoothly across edges, thus, justifying the use of a line-graph in such a case.

Unlike the line-graph Laplacian, the Edge Laplacian captures a notion of flow conservation, which implies that smooth flows should by cyclic~\cite{Schaub2018a}.
To see this, it is insightful to inspect the quadratic regularizer induced by $\bL_e = \bB_1^\top\bB_1$.
Note that this quadratic form can be written as $\bef^\top \bL_e\bef = \|\bB_1 \bef\|_2^2$.
This is precisely the (summed) squared divergence of the flow signal $\bef$, as each entry $(\bB_1 \bef)_i$ corresponds to the difference of the inflow and outflow to node $i$
\begin{align}
    (\bB_1 \bef)_i = \sum_{r \in \{(j,i) | \{i,j\} \in \mathcal{E}, j<i\}} f_r - \sum_{r \in \{(i,j) | \{i,j\}\in \mathcal{E}, i<j \}} f_r,
\end{align}
where $f_r$ is the flow on edge $r=(i,j)$, and we have used a reference orientation induced by the lexicographic order.
As a consequence, all cyclic flows will induce zero cost for the regularizer $\bef^\top \bL_e\bef$, which may also be seen from the fact that $\text{ker}(\bB_1)$ is precisely the cycle space of a graph with incidence matrix $\bB_1$.
Stated differently, any flow that is \emph{not} divergence free, i.e., not cyclic, will be penalized by the quadratic form.
Since by the fundamental theorem of linear algebra $\ker(\bB_1) \perp \text{im} (\bB_1^\top)$, any such non-cyclic flows can be written as a gradient flow $\bef_\text{grad} =\bB_1^\top \mathbf{v}$ for some vector $\mathbf{v}$ of scalar node potentials --- in line with the Hodge decomposition discussed in~\eqref{eq:Hodge_decomposition}.

In contrast to the Edge Laplacian, the full Hodge Laplacian $\bL_1$ includes the additional regularization term $\bef^\top \bB_2\bB_2^\top \bef = \|\bB_2^\top\bef\|_2^2$, which may induce a non-zero cost even for certain cyclic flows. 
More precisely, any cyclic flow that can be written as a curl flow $\bef_\text{curl}=\bB_2\mathbf{t}$, for some vector $\mathbf{t}$ will have a non-zero penalty.
This penalty is incurred despite the fact that $\bef_\text{curl}$ is a cyclic flow by construction (since $\bB_1\bef_\text{curl} = \bB_1\bB_2\mathbf{c} = 0$, the vector $\bef_\text{curl}$ is clearly in the cycle space; see also discussion in Section~\ref{subsection:Hodge_Laplacian}).
The additional regularization term $\|\bB_2^\top\bef\|_2^2$ may thus be interpreted as squared curl penalty.

From a signal processing perspective, the $\bL_1$ based filter thus allows for a more refined notion of a smooth signal. 
Unlike in the Edge Laplacian filter, we do not declare all cyclic flows to be maximally smooth and consist only of frequency (eigenvalue) $0$ basis signals.
Instead a signal can have a high-frequency even if it is cyclic, if it has a high curl component.
Hence, by constructing simplicial complexes with appropriate (triangular) $2$-simplices, we have additional modeling flexibility for shaping the frequency response of an edge-flow filter~\cite{yang2021finite}.

In our example above, this more refined notion of a smooth signal is precisely what leads to an improvement in the filtering performance, since the ground truth signal is a harmonic function with respect to the simplicial complex and thus does not contain any curl components.
We remark that the eigenvector basis of $\bL_e$ can always be chosen to be identical to the eigenvectors of $\bL_1$; thus, we may represent any signal in exactly the same way in a basis of $\bL_e$ or $\bL_1$; however, the frequencies associated with all cyclic vectors will be $0$ for the Edge Laplacian, while there will be cyclic flows with nonzero frequencies for $\bL_1$, in general.
This emphasizes that the construction of faces is an important modeling choice for the selection of an appropriate notion of a smooth signal.

\subsection{Interpolation and semi-supervised learning}
Let us now focus on the interpolation problem for edge-data on a simplicial complex~\cite{Jia2019}.
Analogously to  node signals, we are given a simplicial complex (or its graph skeleton) and a set of ``labeled'' oriented edges $\mathcal{E}^L \subset \mathcal{E}$, i.e., we assume that we have measured the edge-signals on some edges but not on all.
Our goal is now to predict the signals on the unlabeled or unmeasured edges in the set $\mathcal{E}^U \equiv \mathcal{E}\backslash \mathcal{E}^L$, whose cardinality we will denote by $E_U$. 
Following~\cite{Jia2019}, we will again start by considering the problem setup with no $2$-simplices first ($\bB_2=0$), before we consider the general case in which $2$-simplices are present.

To arrive at a well-defined problem for imputing the remaining edge-flows, we need to make an assumption about the structure of the true signal.
Following our above discussions, we will again assume that the true signal has a low-pass characteristic in the sense of the Hodge $1$-Laplacian, i.e., that the edge flows are mostly conserved. 
Let $\bfh$ denote the vector of the true (partly measured) edge-flow.
As discussed in the context of flow smoothing, a convenient loss function to promote flow conservation is the sum-of-squares vertex divergence 
\begin{align}
    \norm{\bB_1 \bfh}^2_2 = \bfh^\top \bB_1^\top \bB_1 \bfh = \bfh^\top \bL_e \bfh.
\end{align}
We can then formalize the flow interpolation problem via the following optimization program
\begin{align} \label{eq:SC_SSL}
        \min_{\bfh} \norm{\bB_1 \bfh}_2^2 + \alpha^2 \cdot \norm{\bfh}_2^2 \text{ s.t. } \hat{f}_r = f_r, \text{ for all measured edges } r \in \mathcal{E}^L,
\end{align}
Note that, in contrast to the node signal interpolation problem, we have to add an additional regularization term $\|\bfh \|_2^2$ to guarantee the uniqueness of the optimal solution.
The reason is that, if there is more than one independent cycle in the network for which we have no measurement available, we may add any cyclic flow on such a cycle while not changing the cost function.
To remedy this aspect, we simply add a $2$-norm regularization which promotes small edge-flow magnitudes by default.
Other regularization terms are possible as well, however this formulation enables us to rewrite the above problem in a least squares form as described below.

To arrive at a least-squares formulation, we consider a trivial feasible solution $\bfh^0$ for~\eqref{eq:SC_SSL} that satisfies $\hat{f}^0_r = f_r$ if $r \in \mathcal{E}^L$ and $\hat{f}^0_r = 0$ otherwise. 
Let us now define the expansion operator $\bm \Phi$ as the linear map from $\mathbb{R}^{E_U}$ to $\mathbb{R}^{E}$ such that the true flow $\bef$ can be written as $\bef = \bfh^0 + \bm\Phi \bef^U$, where $\bef^{U}\in \mathbbm{R}^{E_U}$ is the vector of the unmeasured true edge-flows.
Reducing the number of variables considered in this way, we can convert the constrained optimization problem \eqref{eq:SC_SSL} into the following equivalent unconstrained least-squares estimation problem for the \emph{unmeasured} edges $\bfh^{U}$:
\begin{align}\label{eq:SSL_Le}
    \bfh^{U*} = \argmin_{\bfh^U} \norm{ \left[\begin{array}{c}  \bB_1 \bm \Phi\\ \alpha \bI \end{array} \right] \bfh^U - \left[\begin{array}{c}  -\bB_1\bef^0\\ 0 \end{array} \right] }^2_2.
\end{align}
We illustrate the above procedure by the following example.
\begin{example}
We consider the network structure in Figure \ref{fig:SC_example}A. The ground truth signal is $\bef = (-2,-2,4,-2,3,-7,7,3,4,-4)^{\top}$. We pick five labeled edges at random (colored in Figure \ref{SSL_edge_flow}A). The goal is to predict the labels of the unlabeled edges (in grey with a question mark in Figure \ref{SSL_edge_flow}A). The set of labeled edges is $\mathcal{E}^L = \{(1,3),(1,4),(3,6),(4,5),(5,6)\}$. The set of unlabeled edges is $\mathcal{E}^U = \{(1,2),(2,3),(3,4),(5,7),(6,7)\}$. Solving the optimization program \eqref{eq:SSL_Le}, we obtain the predicted signal $\mathbf{f}^*_{SSL}$ in Figure \ref{SSL_edge_flow}B.
Numerical values are given in Figure \ref{SSL_edge_flow}C.
The Pearson correlation coefficient between $\bef$ and $\mathbf{f}^*_{SSL}$ is 0.99. The $2$-norm of the error is 0.064.

\begin{figure}
    \centering
    \includegraphics[width = \textwidth]{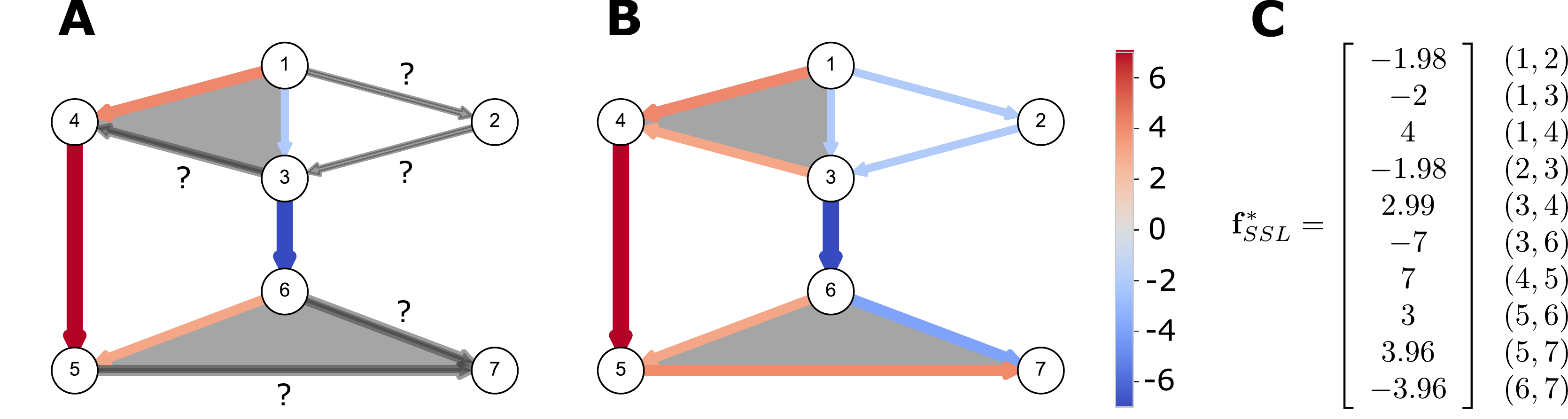}
    \caption{\textbf{Semi-supervised learning for edge flow.} \textbf{A} Synthetic flow. 50\% of the edges are labeled. Labeled edges are colored based on the value of their flow. The remaining edges in grey are inferred from the procedure explained in the text. \textbf{B} Edge flow obtained after applying the semi-supervised algorithm in~\eqref{eq:SSL_Le}. \textbf{C} Numerical value of the inferred signal.}
    \label{SSL_edge_flow}
\end{figure}
\end{example}

Analogously to our discussion above, it may be relevant to include $2$-simplices for the signal interpolation problem.
We interpret such an inclusion of $2$-simplices in two ways.
From the point of view of the cost function, it implies that instead of penalizing primarily gradient flows (which have nonzero divergence), we in addition penalize certain cyclic flows, namely those that have a nonzero curl component.
From a signal processing point of view, it means that we are changing what we consider a smooth (low-pass) signal, by adjusting the frequency representation of certain flows.
Accordingly, one possible formulation of the signal interpolation problem, including information about $2$ simplices is
\begin{align}
\bfh^{\star} &= \argmin_{\bfh} \norm{\bB_1 \bfh}_2^2 + \norm{\bB_2^\top \bfh}_2^2 + \alpha^2 \norm{\bfh}_2^2,
\end{align}
subject to the constraint that the components of $\bfh$ corresponding to measured flows are identical to those measurements.
As in \eqref{eq:SSL_Le}, we can convert this program into the following least-squares problem
\begin{align}
    \bfh^{U \star} &= \argmin_{\bfh^U} \norm{ \left[\begin{array}{c}  \bB_1 \bm\Phi\\ \alpha \bI \\ \bB_2^\top \bm\Phi \end{array} \right] \bfh^U - \left[\begin{array}{c}  -\bB_1 \bef^0 \\ 0 \\ - \bB_2^\top \bef^0\end{array} \right] }^2_2.
\end{align}

\begin{remark}
    Note that the problem of flow interpolation is tightly coupled to the issue of signal reconstruction from sampled measurements.
    Indeed, if we knew that the edge signal to be recovered was exactly bandlimited~\cite{Barbarossa2020}, then we could reconstruct the edge-signal if we had chosen the edges to be sampled appropriately.
    Just like the interpolation problem considered here may be seen as a semi-supervised learning problem for edge labels, finding and choosing such optimal edges to be sampled may be seen as an active learning problem in the context of machine learning .
    While we do not expand further in this tutorial on the choice of edges to be sampled, we point the interested reader to two heuristic active learning algorithms for edge flows presented in \cite{Jia2019}. 
    We also refer the reader to \cite{Barbarossa2018,Barbarossa2020} for a theory of sampling and reconstruction of bandlimited signals on simplicial complexes, and to \cite{barbarossa2020spmag} for a similar overview that includes an approach for \emph{topology inference} based on signals supported on simplicial complexes.
\end{remark}

\subsection{Beyond linear filters: Simplicial neural networks and Hodge theory}\label{section:GNN}

As discussed in Section~\ref{section:gnn-background}, graph neural networks incorporate nonlinear activation functions in the graph signal processing pipeline in order to learn rich representations for graphs.
In order to generalize these architectures to operate on simplicial complexes, we discuss central concepts underpinning graph neural network architectures in order to understand desirable properties of neural networks for higher-order data.
Graph neural networks in the nodal domain typically have two important features in common:
\begin{description}
    \item[Permutation equivariance.]
    Although the nodes are given labels and an ordering for notational convenience, graph neural networks are not dependent on the chosen labeling of the nodes.
    That is, if the node and corresponding input labels were permuted in some way, the output of the graph neural network, modulo said permutation, will not change.
    \item[Locality.]
    Graph neural networks in their most basic form operate \emph{locally} in the graph structure.
    Typically, at each layer a node's representation is affected only by its own state and the state of its immediate neighbors.
    Forcing operations to occur locally is how the underlying graph structure is used to regularize the functional form of the graph neural network.
\end{description}

Based on these two principles, many architectures have been proposed, such as the popular \emph{graph convolutional network} \cite{kipf2016semi}, which mixes one-step graph filters and nodewise nonlinearities for semi-supervised learning on the nodes of a graph.
Indeed, there has been significant study in understanding the nature of graph convolutional architectures in terms of the spectral properties of the chosen shift or filter operation \cite{gama2020stability}.

\subsubsection*{Simplicial Graph Neural Networks}

Motivated by work on graph neural networks in the node space, and the effectiveness of the Hodge Laplacian for representing certain types of data supported on simplicial complexes as in Section~\ref{section:SC}, we now discuss considerations and limitations for building graph neural network architectures based on representations grounded in combinatorial Hodge theory.
This approach to processing data on simplicial complexes generated a flurry of interest recently, with convolutional architectures based on the Hodge Laplacians and boundary maps being proposed in \cite{Ebli:2020,Bunch:2020,glaze2021principled,bodnar2021weisfeiler}.
As before, let $\mathcal{X}$ be a simplicial complex over a finite set of vertices $\mathcal{V}$, with boundary operators $\{\mathbf{B}_k\}_{k=1}^K$, where $K$ is the order of $\mathcal{X}$.

We consider architectures built on the composition of matrix multiplication with boundary operators and/or Hodge Laplacians of varying order, aggregation functions, and nonlinear activation functions that obey \emph{permutation invariance}, \emph{locality}, and the additional properties of \emph{orientation invariance} and \emph{simplicial locality}.

We begin by defining orientation equivariance, which describes a similar property to permutation invariance for graph neural networks~\cite{Roddenberry2019}.
\begin{description}
    \item[Orientation equivariance.]
    If the chosen arbitrary reference orientation of the simplices in $\mathcal{X}$ is changed, the output of the neural network architecture remains the same, modulo said change in orientation.
\end{description}
Due to the arbitrary nature of the simplex orientations, orientation invariance is clearly a desirable property for a neural network architecture to have.
For a simple class of convolutional neural networks for flows, we must choose the nonlinear activation function carefully in order to satisfy this property.
If one were to construct a simple architecture with weight matrices $\mathbf{W}_1,\mathbf{W}_2$ for flows on a simplicial complex based on $\mathbf{L}_1$ of the form
\begin{equation}
    g_{\mathbf{L}_1,\mathbf{W}}(\mathbf{f}) = \sigma\left(\mathbf{L}_1\sigma\left(\mathbf{L}_1\mathbf{f}\mathbf{W}_1\right)\mathbf{W}_2\right),
\end{equation}
we want $g$ to not change when a different orientation is chosen.
Let $\bTheta\in\mathbb{R}^{E\times E}$ be a matrix taking values $\pm 1$ on the diagonal and zeros elsewhere, representing a change in orientation for each edge.
Then, for a flow $\mathbf{f}$ and Hodge Laplacian $\mathbf{L}_1$, this change in orientation is realized by $\bTheta\mathbf{f}$ and $\bTheta\mathbf{L}_1\bTheta$.
Therefore, for orientation equivariance to hold, we need
\begin{equation}\label{eq:or-equivariance}
    g_{\bTheta\mathbf{L}_1\bTheta,\bW}(\bTheta\mathbf{f}) = 
    \bTheta g_{\mathbf{L}_1,\bW}(\mathbf{f})
\end{equation}
to hold for all flows $\mathbf{f}$.
For this to be true, $\sigma$ must be an odd function so that it commutes with $\bTheta$.
A natural extension to the notion of orientation equivariance is \emph{orientation invariance}, which rewrites \eqref{eq:or-equivariance} as
\begin{equation}
    g_{\bTheta\mathbf{L}_1\bTheta,\bW}(\bTheta\mathbf{f}) = 
    g_{\mathbf{L}_1,\bW}(\mathbf{f}).
\end{equation}
This property has greater utility for tasks such as graph classification, where a global descriptor is desired, rather than output on each simplex.

Another consideration that does not typically arise in the design of graph neural networks is data supported on different levels of the graph.
Data on a simplicial complex can lie on, e.g., nodes, edges, and faces \emph{simultaneously}, motivating the need for architectures that pass data along the many levels of a simplicial complex.
Analogous to the property of locality for graph neural networks, we consider a notion of locality for different levels of a simplicial complex.

\begin{description}
    \item[Simplicial locality.]
    At each layer of an architecture with simplicial locality, information exchange only occurs between adjacent levels of the underlying simplicial complex, i.e., the output of a layer restricted to $k-$simplices is dependent only on the input of that layer restricted to $k-1,k,k+1-$simplices.
\end{description}

As an illustrative example, loosely based on the architecture proposed in \cite{Bunch:2020}, consider a small two-layer neural network simultaneously operating over a simplicial complex of nodes, edges, and triangles.
That is, the input to the neural network is a tuple of signals $(\mathbf{v}_0,\mathbf{f}_0,\mathbf{t}_0)$ on the vertices (graph signals), edges (flows), and triangles, respectively, and each layer performs the following computation:
\begin{align}
    \mathbf{v}_k &= \sigma(\mathbf{L}_0\mathbf{v}_{k-1}+\mathbf{B}_1\mathbf{f}_{k-1}) \\
    \mathbf{f}_k &= \sigma(\mathbf{L}_1\mathbf{f}_{k-1}+\mathbf{B}_2\mathbf{t}_{k-1}+\mathbf{B}_1^\top\mathbf{v}_{k-1}) \\
    \mathbf{t}_k &= \sigma(\mathbf{L}_2\mathbf{t}_{k-1}+\mathbf{B}_2^\top\mathbf{f}_{k-1}),
\end{align}
for some odd elementwise activation function $\sigma$.
That is, at each layer, signals on each level of the simplicial complex are either lifted to the next highest level via the coboundary operator, projected to its boundary using the boundary operator, or diffused via the Hodge Laplacian.
This ``lifting'' and ``projecting'' can only occur between adjacent levels of the simplicial complex, due to the fact that the composition of boundary operators is null, \emph{thereby satisfying simplicial locality.}

We now examine the tuple of signals $(\mathbf{v}_2,\mathbf{f}_2,\mathbf{t}_2)$.
First, suppose $\sigma$ is the identity mapping, so that each signal in  $(\mathbf{v}_2,\mathbf{f}_2,\mathbf{t}_2)$ is a linear function of  $(\mathbf{v}_0,\mathbf{f}_0,\mathbf{t}_0)$.
Then, one can check that
\begin{align}
    \mathbf{v}_2 &= 2\mathbf{L}_0\mathbf{B}_1\mathbf{f}_0
    + \mathbf{L}_0(\mathbf{L}_0+\mathbf{I})\mathbf{v}_0 \\
    \mathbf{f}_2 &= (\mathbf{L}_1^2\mathbf{B}_2+\mathbf{B}_2\mathbf{L}_2)\mathbf{t}_0
    +\mathbf{L}_1(\mathbf{L}_1+\mathbf{I})\mathbf{f}_0
    +(\mathbf{L}_1^2+\mathbf{B}_1^\top\mathbf{B}_1)\mathbf{B}_1^\top\mathbf{v}_0 \\
    \mathbf{t}_2 &= \mathbf{L}_2(\mathbf{L}_2+\mathbf{I})\mathbf{t}_0
    +(\mathbf{L}_2\mathbf{B}_2^\top+\mathbf{B}_2^\top\mathbf{L}_1)\mathbf{f}_0.
\end{align}
Notice that each signal is strictly a function of the signals above and below it, even after multiple layers of the architecture are evaluated.
This indicates that our architecture is incapable of incorporating information from nonadjacent levels of the simplicial complex, due to the composition of boundary operators being null: note that similar properties hold for linear variants of this example making use of boundary operators in this way.

This is not the case, though, when $\sigma$ is nonlinear.
While $\mathbf{B}_1\mathbf{B}_2\mathbf{t}=0$ may hold for all signals $\mathbf{t}$ on the faces, $\mathbf{B}_1\sigma(\mathbf{B}_2\mathbf{t})\neq 0$, in general.
By incorporating nonlinear activation functions, we facilitate full incorporation of signals from all levels of the simplicial complex in the output at each level.
We call this property \emph{extended simplicial locality}.
\begin{description}
    \item[Extended simplicial locality.]
    For an architecture with extended simplicial locality, the output restricted to $k-$simplices is dependent on the input restricted to simplices at all levels, not just those of order $k-1,k,k+1$.
\end{description}
Notice that while simplicial locality is defined for each layer of an architecture, extended simplicial locality is a global property, so that both are simultaneously attainable.
There is a trade-off in achieving extended simplicial locality by interleaving nonlinearities: although there is full influence of the entire simplicial structure on all levels of the output, the structure endowed by the boundary operators (namely, the composition of boundary operators being null) is no longer in effect.
Although the Hodge decomposition~\eqref{eq:Hodge_decomposition} can still be applied to the output signals of such an architecture, the expression of the space of $k$-simplex signals strictly in terms of upper and lower incidence through $k-1$ and $k+1$ simplices ceases to hold when considering the input and output jointly, as opposed to linear filters of the Hodge Laplacian.
This motivates further considerations of how nonlinearities may be necessary in modeling higher-order data, such as in the work of \cite{Neuhauser:2020,neuhauser2020multibody}, where it is shown that higher-order opinion dynamics \emph{must} be nonlinear, lest they be equivalently modeled by a purely pairwise system.
That is, we must relax the structure of simplicial complexes in order to represent more general high-order interactions.
In doing this, we exchange the connection to algebraic topology for greater flexibility in modeling.
This naturally leads to the consideration of hypergraphs and associated signal processing ideas, as discussed in the next section.

\section{Modeling higher-order interactions via hypergraphs}\label{section:hypergraphs}

In this section, we discuss \emph{hypergraphs} as an alternative to simplicial complexes to model higher-order analogs of graphs, and then discuss how we can construct appropriate matrix-based and tensor-based shift operators for such hypergraphs to enable the development of signal processing tools.

An important feature of simplicial complexes is that for every simplex present all of its faces are also included in the complex (and recursively the corresponding faces, and so on).
This inclusion property gives rise to the hierarchy of boundary operators, which anchor simplicial complexes to algebraic topology.
However, this subset inclusion property may be an undesirable restriction, if we want to represent interactions that are exclusive to multiple nodes and do not imply the interaction between all the subsets of nodes.
A related problem is the issue of (extended) simplicial locality as discussed in the previous section, which arises from the restrictions imposed on the boundary operators of simplicial complexes.
Finally, while simplices are endowed with a reference orientation and may be weighted, we might be interested in encoding other types of directionality or heterogeneous weighting schemes of group interactions, which are not easily compatible with the mathematical structure of simplicial complexes.

To illustrate the utility of hypergraphs as modelling tools, let us consider a number of concrete examples in which a hypergraph model may be preferred over a simplicial complex, before providing a more mathematical definition.

\begin{example}\label{Ex:hypergraphs}
In a co-authorship network~\cite{han2009understanding}, having a paper with three or more authors does not imply that these people have written papers in pairs. 
Hypergraphs can distinguish these two cases while graphs and simplicial complexes cannot, in general.
Moreover, the relative contribution of the authors to a paper may be different and we thus may want to have a representation that enables us to assign heterogeneous weights within group interactions.
This again can be done using hypergraphs~\cite{inh1}.
An email network may be described using a directed hypergraph~\cite{park2009anomaly}, whenever there exist emails containing multiple senders or multiple receivers. 
This kind of directional information will be difficult to encode in a simplicial complex (while graphs can encode the directionality here, they lose the higher-order information).
Further examples in which hypergraphs appear naturally include word-document networks in text mining~\cite{inh2, zhu2021co}, gene-disease networks in bioinformatics~\cite{goh2007human,aksoy2020hypernetwork}, and consumer-product networks in e-commerce~\cite{li2018tail}.
\end{example}

Mathematically, a typical hypergraph ${\mathcal{H}=(\mathcal{V},\mathcal{E},\omega)}$ consists of a set of vertices $\mathcal{V}$, a set of hyperedges $\mathcal{E}$, and a function $\omega:\mathcal{E}\to\mathbb{R}_{+}$ that assigns positive weights to hyperedges.
Hyperedges generalize edges in the sense that each hyperedge can connect more than two vertices.
In the most common case, where there is one type of node and one type of hyperedge  (namely all hyperedges representing the same type of relationship such as co-authorship), a hypergraph is called homogeneous. 
A hypergraph is called $k$-uniform if all of its hyperedges have the same cardinality $k$. 
Notice, in particular, that a hypergraph is a bona fide generalization of a graph, since a $2$-uniform hypergraph reduces to a graph. 
More interestingly, a simplicial complex may be seen as a hypergraph satisfying the property that every subset of a hyperedge is also a hyperedge. 
Similar to a standard graph, a hypergraph can also be directed in which case each (directed) hyperedge $e$ is an ordered pair $(T(e),H(e))$ where $T(e)$ and $H(e)$ are two disjoint subsets of vertices respectively called the tail and the head of $e$ \cite{directed_hypergraph}.
This flexibility is of interest, e.g., when modelling multiway communication  patterns as illustrated in the example of email networks above.

While the standard framework of hypergraphs is already very flexible, in recent years several more elaborate hypergraph models have been proposed to better represent real-world datasets: 
\begin{enumerate}[(1)]
\item
Heterogeneous hypergraphs refer to hypergraphs containing different types of vertices and/or different types of hyperedges \cite{hg1,hg2,hg3,hg4} and may thus be seen as a generalization of multilayer and multiplex networks.
For example, in a GPS network \cite{gps}, a hyperedge can have three types of vertices (user, location, activity). 
Another example is online social networks such as Twitter, in which we can have different types of vertices including users, tweets, usertags, hashtags and groups as well as multiple types of hyperedges such as `users release tweets containing hashtags or not', `users join groups', and `users assign usertags to themselves'~\cite{li2013link}.
\item
Edge-dependent vertex weights are introduced into hypergraphs in \cite{inh1,inh2,zhu2021co} to reflect the different contribution (e.g., importance or influence) of vertices in the same hyperedge. 
More precisely, for each hyperedge $e\in\mathcal{E}$, a function $\gamma_e:e\to\mathbb{R}_{+}$ is defined to assign positive weights to vertices in this hyperedge. 
For instance, in the co-authorship network in Example~\ref{Ex:hypergraphs}, the different levels of contribution of the authors of a paper can be encoded as edge-dependent vertex weights. 
If $\gamma_e(v)=\gamma_{e'}(v)$ for every vertex $v$ and every pair of hyperedges $e$ and $e'$ containing $v$, then we say that the vertex weights are edge-independent. 
Such hypergraphs are also called vertex-weighted hypergraphs \cite{vertex_weighted}. 
Moreover, if $\gamma_e(v)=1$ for all vertices $v$ and incident hyperedges $e$, the vertex weights are trivial and we recover the homogeneous hypergraph model. 
\item 
In order to leverage the fact that different subsets of vertices in one hyperedge may have different structural importance, the concept of an inhomogeneous hyperedge is proposed in \cite{panli1}. Each inhomogeneous hyperedge $e$ is associated with a function $w_e:2^e\to\mathbb{R}_{\geq 0}$ that assigns non-negative costs to different cuts of the hyperedge, where $2^e$ denotes the power set of $e$.
The weight $w_e(\mathcal{S})$ indicates the cost of partitioning the hyperedge $e$ into two subsets $\mathcal{S}$ and $e\setminus\mathcal{S}$. 
This is called a submodular hypergraph when $w_e$ satisfies submodularity constraints~\cite{panli2}.
\end{enumerate}

Similar to graphs and simplicial complexes, a key factor for developing signal processing tools for hypergraphs is the definition of an appropriate shift operator.
For simplicial complexes, we argued that the Hodge Laplacian is a natural and principled operator for this purpose.
For hypergraphs there are two major approaches to their mathematical representation, which induce different kinds of shift operators.

The first option is to use a matrix-based representation and derive a shift operator from it, akin to the approach of GSP.
As any matrix may be interpreted as an adjacency matrix of a graph and thus induces a weighted, directed graph, this procedure may be understood as first deriving a graph-based representation of the hypergraph and then using an algebraic representation of this graph (e.g., adjacency or Laplacian matrices) as the algebraic shift operator of the hypergraph.

The second option is to represent the hypergraph using a tensor, i.e., a multi-dimensional array representation instead of the 2-dimensional array representation provided by matrices (we refer to~\cite{kolda2009tensor, cichocki2015tensor, sidiropoulos2017tensor} for a general introduction to tensors and tensor decompositions).
While this provides, in principle, a richer set of possible representations of the shift operator, there are also challenges associated with this procedure as the definition of a hypergraph signal and its processing is less grounded in GSP and related techniques. 
In the following subsections, we respectively discuss these two choices of representations, starting with matrix-based representations.

\subsection{Matrix-based hypergraph representations}\label{SS:mat-based}
\begin{figure}
    \centering
    \includegraphics[scale=0.3]{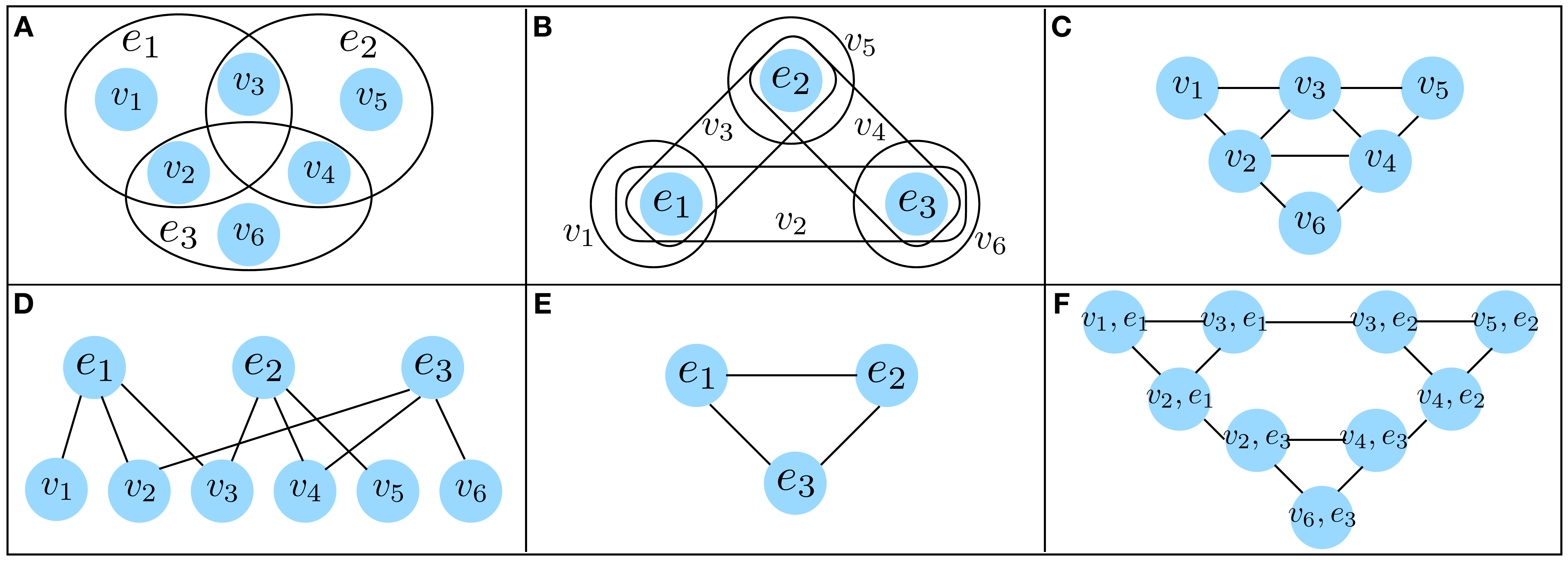}
    \caption{\textbf{Different transformations on an example hypergraph.} \textbf{A} The original hypergraph. \textbf{B} The dual hypergraph. \textbf{C} The clique expansion. \textbf{D} The star expansion. \textbf{E} The line graph. \textbf{F} The line expansion.}
    \label{fig:hg_proj}
\end{figure}

The most common approach to deal with hypergraph-structured data is to encode the hypergraph as a matrix.
When interpreting the corresponding matrices as graphs, many of these matrix-based approaches can thus, alternatively, be viewed as deriving a graph representation for the hypergraph.
Accordingly, these approaches are often described in terms of graph expansions. 
We prefer the term matrix representation here, as the fact that we encode a particular data structure via a matrix does not imply that the data structure is itself a graph (possibly with weights and signed edges).
For instance, we studied matrix-based representations of simplicial complexes in the previous sections, but this would typically not be considered a graph expansion of a simplicial complex.

Let us now discuss some of the most common matrix-based hypergraph representations and transformations (see Figure~\ref{fig:hg_proj} for a visual overview of the discussed variants), including the so-called clique and star expansions as the most popular variants~\cite{Agarwal2006}.
To this end, consider a homogeneous hypergraph $\mathcal{H}=(\mathcal{V},\mathcal{E},\omega)$ and define the vertex-to-hyperedge incidence matrix as $\mathbf{Z} \in \mathbbm{R}^{|\mathcal{V}|\times |\mathcal{E}|}$ with entries $Z_{ve} = 1$ if vertex $v$ belongs to hyperedge $e$.
In addition, we will represent the weights of the hyperedges by the diagonal matrix $\bW \in \mathbbm{R}^{|\mathcal E|\times |\mathcal{E}|}$, whose diagonal corresponds to the hyperedge weights.

Let us first consider the so called star-graph expansion (Figure~\ref{fig:hg_proj}D)~\cite{zhou2006learning,sole1996spectra}.
Using the above defined matrices, the star-graph expansion can be explained by constructing the following adjacency matrix $\bA_*$ of a bipartite graph
\begin{equation}
    \bA_* = 
    \begin{bmatrix}
    \mathbf{0} & \mathbf{Z}\bW\\
    \bW \mathbf{Z}^\top & \mathbf{0} 
    \end{bmatrix}
    \in \mathbbm{R}^{(|\mathcal{V}|+|\mathcal{E}|)\times(|\mathcal{V}|+|\mathcal{E}|)}.
\end{equation}
When interpreted in terms of a graph, this construction may be explained as follows: We introduce a new vertex for each hyperedge and each of these vertices is then connected with a weight corresponding to the weight of the hyperedge to all the (original) vertices in this hyperedge.
The constructed weighted graph $\mathcal{G}_*=(\mathcal{V}_*,\mathcal{E}_*, \omega_*)$, thus has a vertex set $\mathcal{V}_* =\mathcal{V}\cup\mathcal{E}$, an edge set $\mathcal{E}_*=\{(v,e):v\in e, e\in\mathcal{E}\}$, and an edge weight function $\omega_*(v,e) = \omega(e)$.
Many other weight functions are possible here as well, e.g., we may normalize by the cardinality of the hyperedges.
By constructing appropriate Laplacian operators (combinatorial or normalized) of such a star expansion matrix, we can thus obtain a shift-operator for the hypergraph in a straightforward fashion.

An alternative matrix-based representation that can be derived from the same matrices defined above is the clique expansion (Figure~\ref{fig:hg_proj}C)~\cite{bolla1993spectra, rodri2002laplacian, rodriguez2003laplacian, gibson2000clustering}. 
In matrix terms, this corresponds to projecting out the hyperedge dimension of the incidence matrix $\mathbf{Z}$.
Specifically, if we assume unit hyperedge weights for simplicity, the clique expansion may be computed by forming the product $\mathbf{ZZ}^\top$.
As this matrix has a nonzero diagonal, we can simply set the diagonal of this matrix to zero to obtain a basic clique expansion matrix $\bA_\text{c} = \mathbf{ZZ}^\top - \text{Diag}(\text{diag}(\mathbf{ZZ}^\top))$.
By including various weighting factors, alternative variants of this matrix can be derived.
The name clique expansion becomes intuitive if we again interpret $\bA_\text{c}$ as the adjacency matrix of a graph: The above construction corresponds to replacing every hyperedge with a clique subgraph.
More precisely, the clique expansion leads to the adjacency matrix of a graph $\mathcal{G}_c=(\mathcal{V}_c,\mathcal{E}_c,\omega_c)$ in which $\mathcal{V}_c=\mathcal{V}$, $\mathcal{E}_c=\{(u,v):u,v\in e, e\in\mathcal{E}, u\neq v\}$. 
One of the most common definitions for the edge weighting function in this context is $\omega_c(u,v)=\sum_{e\in\mathcal{E}:u,v\in e}\omega(e)$, i.e., the edge weight in the graph is simply given by the sum of the weights of hyperedges that contain the two endpoints.
However, many other weighting schemes are conceivable.

As has been shown in~\cite{Agarwal2006}, many hypergraph learning algorithms~\cite{zhou2006learning, sole1996spectra, bolla1993spectra, rodri2002laplacian, rodriguez2003laplacian, gibson2000clustering} correspond to either the clique or star expansions with an appropriate weighting function. 
However, apart from these common expansions, there also exist other methods for projecting hypergraphs to graphs such as constructing a line graph~\cite{bandyopadhyay2020line}. 
This line-graph expansion for hypergraphs (see Figure~\ref{fig:hg_proj}E for an illustration) may be computed in terms of (weighted variants of) the second possible projection of the incidence matrix $\mathbf{Z}$, namely $\mathbf{Z}^\top\mathbf{Z}$.
Apart from these three canonical types of graph representations (star, clique, and line graph) that can be derived from the incidence matrix $\mathbf{Z}$ and additional (weighting) transformations, a few other matrix-based schemes have been proposed for representing hypergraphs.
For instance, the recent paper \cite{yang2020hypergraph} proposes the so-called line expansion of a hypergraph (different from the line graph; see Figure~\ref{fig:hg_proj}F), which is isomorphic to the line graph of its star expansion and aims to unify the clique and star expansions. 
In the line expansion, each incident vertex-hyperedge pair is considered as a ``line node" and two ``line nodes" are connected if they share either the vertex or the hyperedge. 
We would like to remark that in some cases we might be more interested in the dual of one hypergraph in which the roles of vertices and hyperedges are interchanged and the incidence matrix is $\mathbf{Z}^{\top}$~\cite{aksoy2020hypernetwork}; see Figure~\ref{fig:hg_proj}B.

While we have so far considered only homogeneous hypergraphs, Laplacian matrices have also been proposed for more general hypergraph models. 
For instance, \cite{inh1, inh2, panli1} use variants of the clique expansion to derive matrix representations of hypergraphs with edge-dependent vertex weights or inhomogeneous hyperedges. 
Specifically, in \cite{inh1, inh2} hypergraphs with edge-dependent vertex weights are projected onto asymmetric matrices, corresponding to induced directed graphs with self-loops.
The authors then use established combinatorial and normalized Laplacians for digraphs~\cite{chung2005laplacians} applied to these matrices to derive a Laplacian matrix for hypergraphs. 
Finally, in \cite{panli1}, a novel algorithm for assigning edge weights to the graph representation is proposed, allowing for non-uniform expansions of hyperedges. 

As the above discussion shows, there is an enormous variety of matrix-based representations for hypergraphs, and the relative advantages and disadvantages of these constructions are still sparsely understood.
Ultimately, the choice of a particular matrix representation corresponds to a specific model for what constitutes a smooth signal on a hypergraph.
We believe that a better understanding of the spectral properties of the individual constructions will thus be an important step for choosing good matrix representations for different application scenarios.

\subsection{Tensor-based hypergraph representations}

Instead of working with matrix-based representations, hypergraphs can alternatively be represented by tensors.
A tensor is simply a multi-dimensional array, whose order is the number of indices needed to label an element in the tensor~\cite{kolda2009tensor}. 
For instance, a vector and a matrix are a first-order and a second-order tensor, respectively. 
Several different versions of a hypergraph adjacency tensor have been proposed in existing work \cite{michoel2012alignment, ghoshdastidar2017uniform, cooper2012spectra, hu2015laplacian, banerjee2017spectra, ouvrard2017adjacency, pearson2015spectral, benson2015tensor, benson2019three, zhang2019introducing}. 
In this section, we focus on unweighted hypergraphs to keep our exposition accessible and to remain consistent with the majority of the existing work in this domain. 

Due to their relative simplicity, $k$-uniform hypergraphs have been first studied in the literature.
As every hyperedge is of the same order, a $k$-uniform hypergraph with $N$ nodes can be naturally represented by a $k$th-order adjacency tensor $\mathcal{A}\in\mathbb{R}^{N \times N \times \cdots \times N}$, where each index ranges from $1$ to $N$, and the entries of $\mathcal{A}$ are defined as follows~\cite{michoel2012alignment, ghoshdastidar2017uniform}
\begin{equation}\label{E:adjacency_tensor}
{A}_{i_1\cdots i_k} = 
    1, \qquad  \text{ if } \{v_{i_1},\cdots,v_{i_k}\}\in\mathcal{E}.
\end{equation}
Every other entry in $\mathcal{A}$ is set to zero.
Similarly to how it can be meaningful to normalize the adjacency matrix, normalized versions of this adjacency tensor have been proposed as well.
In \cite{cooper2012spectra}, the tensor in~\eqref{E:adjacency_tensor} is normalized by $1/(k-1)!$. 
This normalization guarantees that the degree of a vertex $v_i$, i.e., the number of hyperedges that it belongs to, can be retrieved by summing the entries in the tensor whose first mode index is $i$, namely $\deg(v_i)=\sum_{i_2,\cdots,i_k=1}^N {A}_{ii_2\cdots i_k}$; see~\cite{ouvrard2017adjacency}.
This is desirable because it resembles the way of obtaining the degree of a vertex in a graph from its adjacency matrix.
Another normalized adjacency is proposed in \cite{hu2015laplacian} where
\begin{equation}
{A}_{i_1\cdots i_k} = \frac{1}{(k-1)!}\prod_{j=1}^k \frac{1}{\sqrt[k]{\deg(v_{i_j})}}, \qquad  \text{ if } \{v_{i_1},\cdots,v_{i_k}\}\in\mathcal{E},
\end{equation}
and the rest of the entries are equal to zero.
Its associated normalized Laplacian tensor is defined as $\mathcal{L}=\mathcal{J}-\mathcal{A}$ where $\mathcal{J}$ is a tensor of the same size as $\mathcal{A}$, and its entry ${J}_{ii\cdots i}=1$ if $\deg(v_i)>0$ and $0$ otherwise.
This normalization ensures that $\mathcal{L}$ has certain desirable spectral properties that mimic those of the normalized graph Laplacian~\cite{hu2015laplacian}. 
For example, the eigenvalues of $\mathcal{L}$ as defined in~\cite{qi2005eigenvalues} are guaranteed to be contained in $[0,2]$.
Having a bounded spectrum has shown to be useful in GSP for the stability analysis of graph filters~\cite{isufi2017autoregressive}.

For hypergraphs with non-uniform hyperedges, i.e., hyperedges of different sizes, the above construction does not extend easily.
Since some edges will have smaller cardinality than others, some indices in the adjacency tensor would simply be undefined.
A naive approach would be to keep an adjacency tensor for each observed cardinality of hyperedges, but this approach is computationally impractical.
An alternative is to augment the above construction of an adjacency tensor for general homogeneous hypergraphs as follows.
Denote by $m$ the cardinality of the largest hyperedge across all hyperedges $e\in\mathcal{E}$.
Then, we construct an adjacency tensor of order $m$ according to the following rules~\cite{banerjee2017spectra}.
For every hyperedge $e=\{v_{i_1},\cdots,v_{i_s}\}\in\mathcal{E}$ of cardinality $s\leq m$, we assign the following nonzero entries to $\mathcal{A}$
\begin{equation}\label{E:general_adj_tensor}
{A}_{p_1p_2\cdots p_m} = s \cdot \left(\sum_{l_1,\cdots,l_s\geq 1, \sum_{j=1}^s l_j=m} \frac{m!}{l_1!l_2!\cdots l_s!} \right)^{-1}, 
\end{equation}
where the indices $p_1,p_2,\cdots,p_m$ are chosen in all possible ways from $\{i_1,i_2,\cdots,i_s\}$ such that every element of this latter set is represented at least once.
The rest of the entries of $\mathcal{A}$ are set to zero. 
The Laplacian tensor is then defined as $\mathcal{L}=\mathcal{D}-\mathcal{A}$ where $\mathcal{D}$ is a super-diagonal tensor of the same size as $\mathcal{A}$ and with entries ${D}_{ii\cdots i}$ equal to the degree of vertex $v_i$. 
To illustrate definition \eqref{E:general_adj_tensor}, consider the following example.
\begin{example}
Consider a hypergraph composed of four nodes $v_1,v_2,v_3,v_4$ and two hyperedges $e_1=\{v_1,v_2,v_3\}$ and $e_2=\{v_3,v_4\}$. We have that $m=\max \{|e_1|, |e_2|\} = 3$ and the adjacency tensor is of size $4\times 4\times 4$. For $e_1$, the corresponding $s=|e_1|=3$ and $l_1=l_2=l_3=1$, thus the corresponding entries in the tensor are defined as $A_{123}=A_{132}=A_{213}=A_{231}=A_{312}=A_{321}= 3 / 3! = 1/2$. For $e_2$, the corresponding $s=|e_2|=2$ and there are two choices for $l_1$ and $l_2$, i.e., $l_1=1, l_2=2$ or $l_1=2, l_2=1$.
Thus we have  $A_{344}=A_{434}=A_{443}=A_{334}=A_{343}=A_{433}= 2 \cdot (3!/2! + 3!/2!)^{-1} = 1/3$. 
The remaining entries are set to zero.
\end{example}

Having defined adjacency and Laplacian tensors, we can now construct appropriate shift operators based on these tensors.
In the context in which we are interested in processing signals $\mathbf{y}=[y_1,y_2,\cdots,y_N]^\top$ defined on the nodes, the following approach has been proposed~\cite{zhang2019introducing}.
First, given the signal vector $\by$ construct the following $(m-1)$th-order outer product tensor $\mathcal{Y}\in\mathbbm{R}^{N\times \cdots\times N}$ as
\begin{equation}\label{E:hsignal}
\mathcal{Y} = \underbrace{\mathbf{y} \circ \cdots \circ \mathbf{y} }_{m-1 \text{ times}}, \text{ with entries } Y_{i_1i_2\cdots i_{m-1}} = y_{i_1}y_{i_2}\cdots y_{i_{m-1}},
\end{equation}
where $\circ$ denotes the tensor outer product and $m$ is the order of the adjacency or Laplacian (shift) tensor of interest. 
Then, the hypergraph shift operation leading to the output signal $\by^\text{out} \in \mathbbm{R}^N$ is defined elementwise as
\begin{equation}\label{E:hshift}
 y_i^\text{out} =\sum_{j_1,\cdots,j_{m-1}=1}^N S_{ij_1\cdots j_{m-1}}y_{j_1}y_{j_2}\cdots y_{j_{m-1}},
\end{equation}
where $S_{ij_1\ldots j_{m-1}}$ correspond to the entries of the chosen shift tensor.
Equivalently, we may express the above in terms of tensor products as
\begin{equation}
 \mathbf{y}^\text{out} = \mathcal{S}\mathcal{Y}.
\end{equation}
Note that, due to the symmetry of the tensor $\mathcal{S}$, it does not matter which mode we leave out in the tensor multiplication, i.e., which of the indices is kept fixed to $i$ in~\eqref{E:hshift}.
Furthermore, for the specific case  where $m=2$, we have that $\mathcal{Y}=\mathbf{y}$ in~\eqref{E:hsignal} and the shift operation in~\eqref{E:hshift} boils down to a standard matrix-vector multiplication as in GSP.

\begin{figure}
    \centering
    \includegraphics[scale=0.3]{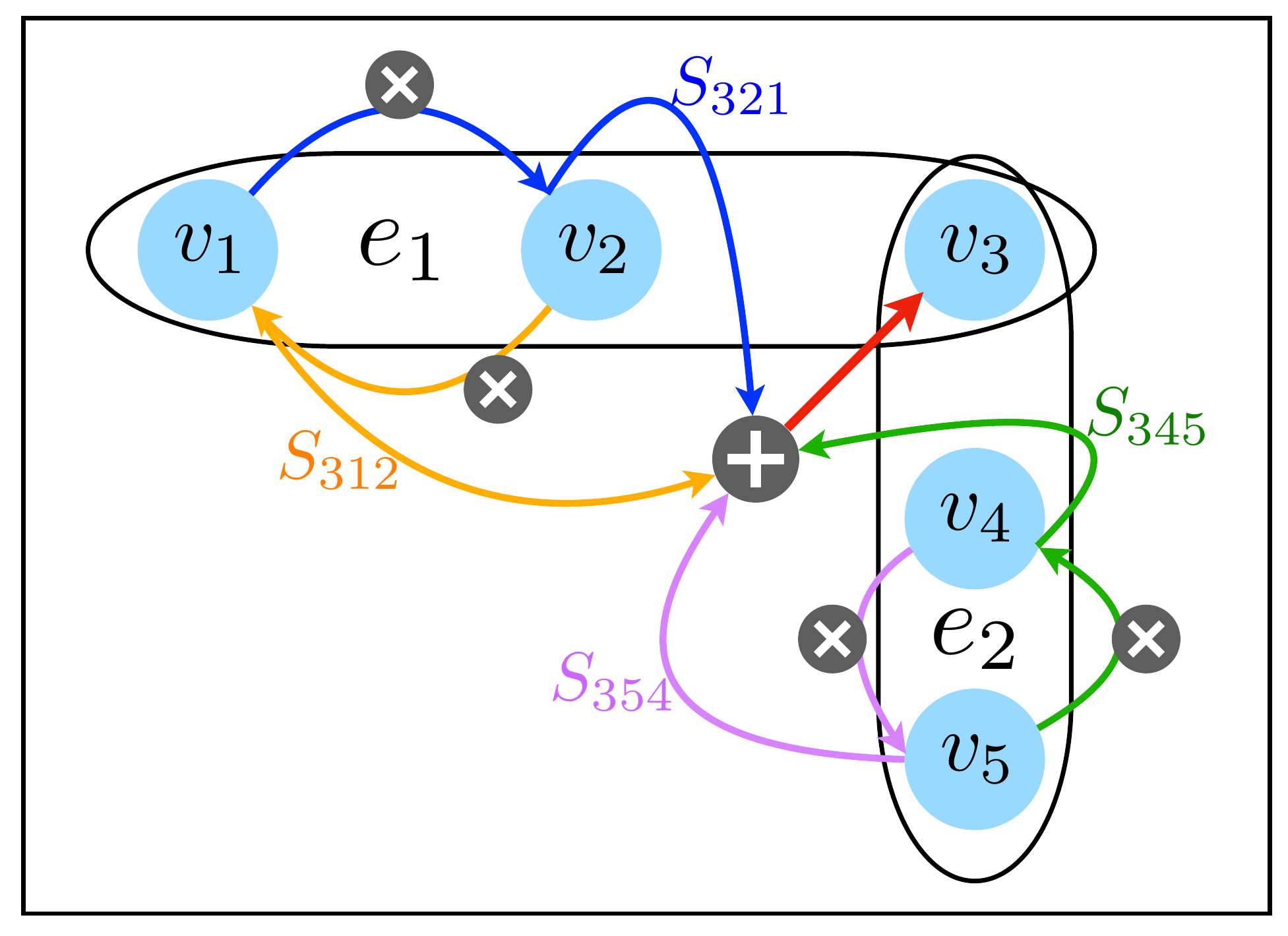}
    \caption{\textbf{Tensor based shift operator on a hypergraph.} The output of $y^{\mathrm{out}}_3$ at vertex $v_3$ is determined by a weighted sum over the hyperedges incident to $v_3$, where the summands correspond to the products of the vertex signals within the respective hyperedges excluding $v_3$. [Figure adapted from Fig. 10(a) in~\cite{zhang2019introducing}]. }
    \label{fig:hg_shift}
\end{figure}

\begin{example}
Consider the hypergraph in Figure~\ref{fig:hg_shift}: vertex $v_3$ is contained in two hyperedges $e_1=\{v_1,v_2,v_3\}$ and $e_2=\{v_3,v_4,v_5\}$. 
We define the adjacency tensor as the shift tensor $\mathcal{S}$.
According to \eqref{E:hshift}, the output signal at vertex $v_3$ after one hypergraph shift is computed as
\begin{equation}\label{E:hg_shift_eg}
y^{\mathrm{out}}_3 = S_{321}\times y_2y_1 + S_{312}\times y_1y_2 + S_{354}\times y_5y_4 + S_{345}\times y_4y_5.
\end{equation}
As in the graph case where the entry $S_{ij}$ of the shift operator indicates the shift from vertex $v_j$ to vertex $v_i$, the entry $S_{ij_1\cdots j_{m-1}}$ of the hypergraph shift operator indicates the shift in one hyperedge following the order $v_{j_{m-1}}\to v_{j_{m-2}}\to \cdots\to v_{j_1}\to v_i$. 
Figure~\ref{fig:hg_shift} illustrates the process defined by~\eqref{E:hg_shift_eg}. 
\end{example}

\subsection{Comparison between matrix-based and tensor-based hypergraph representations} 

The major advantage of matrix-based methods is that a lot of well-developed graph-related algorithms can be directly utilized. 
However, if the resulting matrix representation is akin to a graph in that it only encodes pairwise relations between vertices (clique expansion), or hyperedges (line graphs), there will be some information loss, in general, compared to the original hypergraph structure.
In contrast, for the star-expansion, all the incidence information is kept in the matrix representation.
However, the resulting graph is bipartite. The bipartite graph structure might be undesirable for some applications since there are no explicit links between the same types of vertices and there are much fewer algorithms tailored for bipartite graphs than those for simple graphs~\cite{yang2020hypergraph}.

Compared with matrix representations, tensors can better retain the set-level information contained in hypergraphs. 
However, tensor computations are more complicated and lack algorithmic guarantees \cite{benson2015tensor}. 
For example, determining the rank of a specific tensor is NP-hard \cite{haastad1989tensor}. 
Most existing papers have focused on super-symmetric tensors~\cite{qi2005eigenvalues}, while more general tensors are less explored. 
Indeed, how to best leverage tensor-based representations to study hypergraphs that are not homogeneous is an open problem.

\begin{remark}\label{remark-nonlinearL}
There is a rich and complementary line of research on \emph{nonlinear} Laplacian operators. 
In~\cite{louis2015hypergraph, chan2018spectral}, a continuous diffusion process on the hypergraph is considered to define a Laplacian operator that enables a Cheeger-type inequality for hypergraphs.  
To understand this diffusion process, suppose that, at some instant, there is some signal $\mathbf{y}\in\mathbb{R}^{|\mathcal{V}|}$ defined on the vertices of a hypergraph. 
Each hyperedge $e\in\mathcal{E}$ directs flow from vertices $S_e(\mathbf{y})=\argmax_{v_i\in e}y_i$ having the maximum signal value to vertices $I_e(\mathbf{y})=\argmin_{v_i\in e}y_i$ having the minimum signal value, at a total rate of $c_e=\omega(e)\cdot\max_{v_i,v_j\in e} |y_i-y_j|$.
As the diffusion progresses, the cardinality of $S_e(\mathbf{y})$ and $I_e(\mathbf{y})$ increases, conferring a nonlinear nature to the diffusion process, which can be modeled through a nonlinear Laplacian.
A generalization of this process was proposed in~\cite{chan2020generalizing}, where hyperedges can act as mediators to receive flow from vertices in $S_e(\mathbf{y})$ and deliver flow to those in $I_e(\mathbf{y})$.
Moreover, a unifying framework was recently presented in~\cite{yoshida2019cheeger} by proposing a Cheeger inequality for submodular transformations. 
In particular, the Laplacian operators as well as the Cheeger inequalities for undirected graphs, directed graphs and hypergraphs can be recovered by defining proper submodular transformations; see \cite{yoshida2019cheeger} for more details.
In~\cite{panli2}, similar results have been independently obtained for symmetric submodular transformations. 
\end{remark}

\section{Signal processing and learning on hypergraphs}\label{section:learning_hypergraphs}

Mimicking the respective developments in Section~\ref{ss:graph_signal_processing} for graphs and Section~\ref{section:learning_SC} for simplicial complexes, in this section we consider the four signal processing setups for hypergraphs equipped with the algebraic representations developed in Section~\ref{section:hypergraphs}.

\subsection{Fourier analysis, node and hyperedge embeddings}

As stated in Section \ref{SS:mat-based}, shift operators for hypergraphs can be represented via matrices.
The corresponding eigenvectors may then be used as Fourier modes and, thus, most GSP tools discussed in Section \ref{section:preliminariesGraphs} can be directly translated to hypergraphs for matrix-based hypergraph shift operators. However, unlike for graphs, even an undirected hypergraph may result in an \emph{asymmetric} matrix, e.g., if hyperedge weightings are considered.
Hence, one may have to adopt tools from GSP for directed graphs in this case; see~\cite{marques2020signal} for a more detailed exposition of these issues.

In contrast to matrix-based shift operators, the notion of Fourier analysis for hypergraphs represented via tensors is far less developed.
Nonetheless, we may proceed analogously to the matrix case and define Fourier modes via a tensor decomposition, in lieu of the eigenvector decomposition.
Specifically, we can consider the orthogonal canonical polyadic (CP) decomposition~\cite{afshar2017cp} of the adjacency tensor $\mathcal{A}$ (other representative tensors can also be considered) given by
\begin{equation}\label{E:cp}
\mathcal{A}=\sum_{r=1}^R \lambda_r\cdot \underbrace{\mathbf{v}_r\circ\cdots\circ\mathbf{v}_r}_{m \text{ times}}, 
\end{equation}
where $\lambda_r$ are scalars, and $R$ is the so-called rank of the tensor (i.e., $R$ is the smallest number such that $\mathcal{A}$ can be represented as a weighted sum of rank-1 outer-product tensors).
Using this decomposition, the hypergraph Fourier basis can then be defined as $\mathbf{V}=[\mathbf{v}_1,\cdots, \mathbf{v}_N]$.
When $R<N$, the first $R$ vectors are determined via the CP decomposition and $N-R$ additional vectors satisfying specific conditions are selected to complete the basis (see Section \uppercase\expandafter{\romannumeral3}-F in \cite{zhang2019introducing} for details). 
Similar to the matrix case, the hypergraph Fourier frequencies are defined as the coefficients $\lambda_r$ associated with the rank-1 terms in the decomposition. 

Extending the arguments from the matrix case to the tensor case, the hypergraph Fourier transform (HGFT) and inverse HGFT (iHGFT) \cite{zhang2019introducing} are then defined as
\begin{equation}\label{E:HGFT}
\tilde{\mathbf{y}} = (\mathbf{V}^{\top}\mathbf{y})^{m-1}, \quad \mathbf{y} = \mathbf{V}\tilde{\mathbf{y}}^{\frac{1}{m-1}},
\end{equation}
where $\mathbf{y}^{m}=[y_1^m,\cdots,y_N^m]^{\top}$ denotes the $m$-th power of each entry of $\mathbf{y}$. 
By introducing such definitions, an application of the tensor shift can be equivalently interpreted as a HGFT followed by a combined operation consisting of filtering in the Fourier domain plus iHGFT (cf. Equation (25) in \cite{zhang2019introducing}).
Observe that, when $m=2$, the hypergraph shift defined in \eqref{E:hshift}, and the HGFT and iHGFT defined in \eqref{E:HGFT} have the same form as the corresponding concepts in GSP (cf. Section~\ref{ss:gsp}).

Similar to how graph Fourier modes can be used to derive node embeddings, the same can be done for hypergraphs using either matrix or tensor representations.
For matrix representations, the procedure is entirely analogous.
For tensor representations, we have to use a tensor decomposition but can proceed in a similar fashion once the (tensor-based) Fourier modes are derived.
While tensor-based embeddings have only been scarcely considered in the literature so far, e.g., \cite{sharma2018hyperedge2vec} represents the dual hypergraph (see Figure~\ref{fig:hg_proj}B) using tensors and learns hyperedge embeddings by performing a symmetric tensor decomposition~\cite{comon2008symmetric, kolda2015numerical}. 
Finally, the embedding of \emph{heterogeneous} hypergraphs entails additional challenges that can be (partially) addressed through nonlinear neural network based approaches; see~\cite{hg2} for more details.

\subsection{Signal smoothing and denoising}\label{ss:hg_denoise}

As was the case for graphs and simplicial complexes, one can leverage the structure of hypergraphs to solve inverse problems associated with signals defined on them.
For the specific problem of denoising, the assumption is that the signal to be recovered is smooth in the hypergraph, where smoothness typically encodes the fact that tightly connected nodes should have similar signal values.
For instance, in the co-authorship network in Example~\ref{Ex:hypergraphs}, if two authors share many papers (hyperedges) either written solely by them or in collaboration with others, one would expect a signal that represents research interests to take similar values for the two mentioned authors.
This notion of homophily is well established for graphs and naturally extends to hypergraphs.

Mathematically, in line with the graph case in Section~\ref{ss:graph_denoise}, we assume that we observe a noisy version $\mathbf{y} = \mathbf{y}^0 + \boldsymbol{\epsilon}$ of the true underlying signal defined on the node set of our hypergraph $\mathcal{H}$.
Then, we can try to estimate $\mathbf{y}^0$ by solving the optimization problem
\begin{equation}\label{E:hg_denoise}
\min_{\hat{\mathbf{y}}} \|\hat{\mathbf{y}} - \mathbf{y}\|_2^2 + \alpha \Omega_{\mathcal{H}}(\hat{\mathbf{y}}),
\end{equation}
where the first term is to constrain the denoised signal $\hat{\mathbf{y}}$ to be close to the observation $\mathbf{y}$ and the second term is a regularizer shaped by the structure of $\mathcal{H}$.

A possible choice for $\Omega_{\mathcal{H}}(\hat{\mathbf{y}})$ is to select a Laplacian matrix representation of $\mathcal{H}$ (cf. Section~\ref{SS:mat-based}) and set the regularizer to the quadratic form as in the graph case~\cite{Agarwal2006, zhou2006learning}.
From the discussion after~\eqref{eq:graph_denoising} it follows that the optimal solution $\hat{\mathbf{y}}$ will then be a low-pass version of $\mathbf{y}$ where the bases for low and high frequencies depend on the specific graph expansion selected.
The most common one is to consider the clique expansion, in which we have
\begin{equation}\label{E:hg_denoise_c}
\Omega_{\mathcal{H}}(\hat{\mathbf{y}}) = \sum_{(u,v)\in\mathcal{E}_c} \omega_{c}(u,v)(\hat{y}_u - \hat{y}_v)^2 = \hat{\mathbf{y}}^{\top} \mathbf{L}_{c} \hat{\mathbf{y}},
\end{equation}
where $\mathbf{L}_c$ corresponds to the graph Laplacian obtained via clique expansion of the hypergraph. 
Alternatively, one can rely on tensor-based representations for hypergraphs in the definition of $\Omega_{\mathcal{H}}(\hat{\mathbf{y}})$. 
In particular, we can set the regularizer to be equal to the tensor-based total variation in~\cite{zhang2019introducing}.
In this case, smooth signals would also be promoted but the meaning of a smooth signal will correspond to one that suffers little change under a tensor shift as defined in~\eqref{E:hshift}.

An alternative regularizer based on the Lov\'asz extension of the hypergraph cut has also been proposed~\cite{hein2013total}. More specifically, a parametric family of regularizers was considered
\begin{equation}\label{E:lovasz_regularizer}
\Omega_{\mathcal{H},p}(\hat{\mathbf{y}}) = \sum_{e\in\mathcal{E}} \omega(e) \left( \max_{u\in e} \hat{y}_u - \min_{v\in e} \hat{y}_v \right)^p,
\end{equation}
which can be shown to be convex for $p \geq 1$. 
Consequently, the optimization problem~\eqref{E:hg_denoise} remains convex and, in particular, tailored efficient algorithms have been proposed for $p=1$ and $p=2$; see~\cite{hein2013total}.
In interpreting~\eqref{E:lovasz_regularizer} we can see that $\Omega_{\mathcal{H},p}(\hat{\mathbf{y}})$ induces yet another related notion of smoothness. For every hyperedge $e \in \mathcal{E}$ we look at the difference between the extreme values of the signal attained at the nodes contained in $e$, we scale this penalization by the weight of the hyperedge, and we sum over all hyperedges. 
Intuitively, this regularizer promotes signals that are constant within the hyperedges. 
Moreover, the power $p$ controls the form of the deviations from these piecewise constant signals. For example, the sparsity promoting $p=1$ would encourage the signal variation to be zero within some hyperedges and possibly high in others, whereas $p=2$ would promote a low (possibly non-zero) variation across all hyperedges.

If we consider a general submodular function $F_e$ instead of the hypergraph cut, then \eqref{E:lovasz_regularizer} can be generalized as 
\begin{equation}\label{E:generalized_regularizer}
\Omega_{\mathcal{H},p}(\hat{\mathbf{y}}) = \sum_{e\in\mathcal{E}}[f_e(\hat{\mathbf{y}})]^p,
\end{equation}
where $f_e$ is the Lov\'{a}sz extension of $F_e$ (cf. Remark~\ref{remark-nonlinearL}).
The optimization problem~\eqref{E:hg_denoise} equipped with \eqref{E:generalized_regularizer} are respectively referred to as decomposable submodular function minimization (DSFM) for $p=1$~\cite{stobbe2010efficient,jegelka2013reflection,nishihara2014convergence,ene2015random,ene2017decomposable,li2018revisiting} and quadratic DSFM (QDSFM) for $p=2$~\cite{li2020quadratic}.
Similar to~\eqref{E:hg_denoise_c} which can also be written as $\langle\hat{\mathbf{y}}, \mathbf{L}_c\hat{\mathbf{y}}\rangle$, \eqref{E:generalized_regularizer} can be viewed as $\langle\hat{\mathbf{y}}, \mathcal{L}(\hat{\mathbf{y}})\rangle$ for some Laplacian operator $\mathcal{L}$ depending on $F_e$.

\subsection{Signal interpolation on hypergraphs}
As discussed in the previous sections, signal interpolation and smoothing are closely related problems.
Successful signal interpolation from an observed subset $\mathcal{V}^L$ hinges to a large extent on the selection of a sensible model for a (smooth) ground truth signal that is compatible with the observed (desired) signal characteristics.
For a chosen signal model, we may then again set up an optimization problem for interpolating hypergraph signals as
\begin{equation}\label{E:hg_interpolate}
\min_{\hat{\mathbf{y}}} \Omega_{\mathcal{H}}(\hat{\mathbf{y}}), \quad\text{ s.t. } \hat{y}_v = y_v \text{ for all } v \in \mathcal{V}^L,
\end{equation}
where $\Omega_\mathcal{H}$ is a regularizer chosen to promote the desired signal characteristics, e.g., a low-pass signal.
Like for graphs and simplicial complexes, many choices for the regularization term are possible here and the optimal choice of a regularizer will generally be dependent on the considered application scenario.
For instance, we may choose to use a regularizer based on the clique expansion or some of the other strategies discussed in Section~\ref{ss:hg_denoise}.
Unlike in the graph and simplicial complex setting, however, for hypergraphs we may also consider tensor-based regularizers, which can offer smoothing and interpolation strategies that are not accessible via matrix-based approaches.
Developing and analyzing such approaches for hypergraphs appears to be an interesting avenue for future research.
Problem~\eqref{E:hg_interpolate} can also be converted to another class of optimization problem called submodular Laplacian system~\cite{fujii2018polynomial} which is a generalization of the Laplacian system on graphs~\cite{Zhu2003}.

\subsection{Hypergraph neural networks}\label{ss:hgnn}

The design of neural network architectures to process and learn from data on hypergraphs is a nascent area of research.
Given the developments in graph neural networks mentioned in Section~\ref{section:gnn-background} and the graph expansions for hypergraphs introduced in Section~\ref{SS:mat-based}, an avenue to derive hypergraph neural networks is to compute the graph shifts based on the (clique, star, or line graph) expansions of the hypergraph and then apply a (classical) graph neural network as the one in~\eqref{eq:basic-gcn} or any of the variants surveyed in~\cite{wu2020comprehensive}.

In this direction, one of the earliest hypergraph neural networks~\cite{feng2019hypergraph} adopts the hypergraph Laplacian matrix associated with a weighted clique expansion in~\cite{zhou2006learning} as a graph shift and then implements a graph convolutional network~\cite{defferrard2016convolutional, kipf2016semi} where shift-invariant filters are intertwined with pointwise nonlinearities.
One drawback of the clique expansion is that the resulting graph tends to be dense since a hyperedge is replaced by a number of edges that is quadratic in the size of the hyperedge.
A similar idea is proposed in~\cite{yadati2019hypergcn}, but this convolutional neural network is based on a different hypergraph Laplacian shift (proposed in \cite{chan2020generalizing}), which only requires a linear number of edges for each hyperedge. This provides a more efficient training when compared with that of~\cite{feng2019hypergraph}.
Under this same methodological umbrella, a line hypergraph convolution network is proposed in~\cite{bandyopadhyay2020line}, which expands the hypergraph into a weighted and attributed line graph and then implements a graph convolutional network using the corresponding shift operator.

Architectures grounded on the message-passing variants of graph neural networks (cf. Section~\ref{section:gnn-background}) have also been proposed for hypergraphs.
For instance, in \cite{yang2020hypergraph} the line expansion of the hypergraph is used to define a message passing process where potentially different aggregation functions can be used when passing messages between nodes in the expansion that have either one vertex or one hyperedge in common; see Figure~\ref{fig:hg_proj}-F.
Also, \cite{arya2020hypersage} proposes a generalization of GraphSAGE~\cite{hamilton2017inductive} to hypergraphs, a well-established message passing architecture for graphs.
Recent developments that further extend the state of the art include architectures that tackle the issue that the initially constructed hypergraphs may not be a suitable representation for data~\cite{jiang2019dynamic} 
as well as the formulation of attention~\cite{bai2019hypergraph} and self-attention~\cite{hg3} mechanisms for hypergraphs.

As a closing note, a different perspective is put forth in~\cite{wendler2019powerset}, where a convolutional neural network architecture for powerset data is introduced.
These architectures are designed to learn from set functions, which are signals on the powerset of a given set.
By noticing that cuts in hypergraphs can be interpreted as set functions, these convolutional architectures can be used to solve problems in hypergraphs; see~\cite{wendler2019powerset} for more details.

\section{Discussion}\label{section:Discussion}
Graph signal processing tools have been highly successful in a wide range of applications, ranging from biological to social domains.
This success hinges to a large extent on providing sensible notions for filtering graph signals, such that the relevant dependencies in the signal are kept intact, while undesirable noise components are filtered out.
However, as graphs are only concerned with pairwise relationships, their capabilities for modeling higher-order dependencies are too limited for certain application scenarios in which polyadic relationships are essential.
In such scenarios, simplicial complexes and hypergraphs have recently emerged as two promising conceptual frameworks to address the specific shortcomings of graph-based representations.

Unlike for GSP that can benefit from a rich set of results in spectral graph theory, e.g., to derive appropriate notions of shift operators and signal smoothness, the theory of signal processing on higher-order networks is far less developed.
In this tutorial paper, we provided an introduction to this emerging area, focusing on the choice of appropriate shift operators and associated frequency domain representations, as well as a set of important application scenarios comprising signal smoothing and denoising, signal interpolation, and the construction of nonlinear neural network architectures that can leverage the structure of such higher-order networks.

We believe that this area holds an enormous potential for future developments.
A few relevant future direction include the following.

In the context of simplicial complexes, the investigation of how these should be constructed from data to capture desirable features is certainly one aspect that deserves further research.
As discussed in Sections~\ref{section:SC} and \ref{section:learning_SC}, the choice of appropriate faces has direct consequences on the frequency representation of any signal and is thus highly relevant for applications~\cite{barbarossa2020spmag}.
Similarly, while we discussed only unweighted simplicial complexes for simplicity, the appropriate introduction of weights to emphasize certain features in the data to be investigated is a pertinent issue that should be addressed in future research.
Finally, while we concentrated on simplicial complexes as the most common complexes considered, the restriction to simplicial instead of other type of cell complexes such as cubical complexes is essentially artificial.
From a modeling perspective, simplices may not always capture the appropriate notion of a ``cell'' in a higher-order interaction network. For instance, in traffic and street networks it may be beneficial to consider cubical complexes or other types of models that can better represent the grid-like structure of many of these networks~\cite{ghosh2018topological}.

In the context of hypergraphs, we provide several potential directions for future work. As the first step, constructing a suitable hypergraph is key to the final performance. 
Hence, it is important to develop effective and efficient methods for the construction of hypergraphs from real-world datasets that are usually large-scale.
To better characterize a wider range of datasets, it is necessary to develop more general hypergraph models, such as those considering different types of vertices or having different levels of relations (cf. Section~\ref{section:hypergraphs}). 
A variety of problems that have been well studied in graphs or homogeneous hypergraphs are valuable to be reconsidered and extended to those less explored but more expressive models. 
These problems include, but are not limited to, developing spectral hypergraph theory, node clustering, classification and ranking, link prediction, hypergraph representation learning (especially for heterogeneous hypergraphs in which hyperedges are generally indecomposable~\cite{hg2}), the modeling and analysis of diffusion processes on hypergraphs, tensor-based representations and operations (especially for hypergraphs with edge-dependent vertex weights which are hard to be modeled using super-symmetric tensors), hypergraph kernels, hypergraph classification, and hypergraph alignment.
Although one framework for hypergraph signal processing has already been proposed in \cite{zhang2019introducing}, there are still many open questions. 
In GSP, graph shift and filters can be understood as some network diffusion processes, while it is not clear if and how the hypergraph shift can be connected with a physical process. 
Other problems such as hypergraph filter design, active sampling for reconstruction, and fast hypergraph Fourier transforms are also worth investigating.
Finally, most existing hypergraph neural networks are matrix-based like those introduced in Section~\ref{ss:hgnn}.
A natural extension in this context would be to derive the theory of tensor-based neural networks for hypergraphs.

\section*{Acknowledgements}
This work was partially supported by USA NSF under award CCF-2008555.
MTS acknowledges partial funding from the Ministry of Culture and Science (MKW) of the German State of North Rhine-Westphalia (``NRW R\"uckkehrprogramm''), and the Excellence Strategy of the Federal Government and the L\"ander.
We thank Lucille Calmon for carefully checking and providing feedback on the manuscript.

{\footnotesize
\bibliographystyle{elsarticle-num}
\bibliography{bibliography_signal.bib}}
\end{document}